\documentclass[]{aa}
\usepackage{amsmath}
\usepackage{amssymb}
\usepackage{graphicx}
\usepackage{natbib}
\newcommand{\bl}[1]{\mbox{\boldmath$ #1 $}}

\begin{document}

%\title{Gravitational fragmentation, accretion bursts, and formation of giant %protoplanets on tens-of-au orbits}

\title{Gravitational fragmentation and formation of giant protoplanets on tens-of-au orbits}

% The list of authors, and the short list which is used in the headers.
% If you need two or more lines of authors, add an extra line using \newauthor

\author{Eduard I. Vorobyov\inst{1,2} and Vardan G. Elbakyan\inst{2}}
% List of institutions
\institute{ 
University of Vienna, Department of Astrophysics, Vienna, 1180, Austria \\
\email{eduard.vorobiev@univie.ac.at} 
\and
Research Institute of Physics, Southern Federal University, Roston-on-Don, 344090 Russia
}

% Abstract of the paper

% \abstract{}{}{}{}{} 
% 5 {} token are mandatory
 
\abstract
{}
{Migration of dense gaseous clumps that form in young protostellar disks via gravitational fragmentation is investigated to determine the likelihood of giant planet formation.}
{High-resolution numerical hydrodynamics simulations in the thin-disk limit are employed to compute the formation and long-term evolution of a gravitationally unstable protostellar disk 
around a solar-mass star. }
{We show that gaseous clumps that form in the outer regions of 
the disk (>100~AU) through disk fragmentation are often perturbed by other 
clumps or disk structures, such as spiral arms, and migrate toward the central star on timescales from a few $10^3$ to few $10^4$~yr. The migration timescale is slowest when stellar motion in response to the disk gravity is considered. When approaching the star, 
the clumps first gain mass (up to several tens of $M_{\rm Jup}$), but then quickly lose most of their
diffuse envelopes  through tidal torques. Part of the clump envelope can be accreted on the central star causing an FU-Ori-type 
accretion and luminosity outburst. The tidal mass loss helps the clumps to 
significantly slow down or even halt 
their inward migration at a distance of a few tens of AU from the protostar. The resulting 
clumps are heavily truncated both in mass and size compared to their wider-orbit counterparts, 
keeping only a dense and hot nucleus. During the inward migration, the temperature in 
the clump interiors may exceed the molecular hydrogen dissociation limit (2000~K) and the central 
region of the clump can collapse into a gas giant protoplanet. 
Moreover, migrating clumps may experience close encounters with other clumps, resulting in the ejection of the least massive (planetary-mass) clumps from the disk. We argue that FU-Orionis-type 
luminosity outbursts may be the end product of disk fragmentation and clump inward migration, 
ushering the formation of giant protoplanets on tens-of-au orbits in systems such as HR~8799.}
{}

\authorrunning{Vorobyov \& Elbakyan}
\titlerunning{Formation of giant planets}

\maketitle
%%%%%%%%%%%%%%%%%%%%%%%%%%%%%%%%%%%%%%%%%%%%%%%%%%

%%%%%%%%%%%%%%%%% BODY OF PAPER %%%%%%%%%%%%%%%%%%

\section{Introduction}
Gravitational instability and fragmentation of protostellar disks has long been considered 
as one of the possible formation mechanisms of giant planets and brown dwarfs 
\citep[e.g.,][]{1998Boss,2003RiceArm,2007MayerLufkin,2010VorobyovBasu,2012ZhuHartmann,2013Vorobyov,2015Stamatellos}.  
Disk fragmentation tend to occur at radial distances beyond 100~AU from 
the central star, where the cooling time becomes shorter than the local dynamical timescale
\citep[e.g.,][but see also \citet{2015Meru}]{2003JohnsonGammie,2005Rafikov}.
The importance of disk fragmentation as a likely mechanism for giant planet and brown dwarf 
formation has been reinforced by the 
detection of wide-orbit (from several tens to hundreds AU) planetary and brown dwarf companions to
low-mass stars \citep[e.g.][]{2008MaroisMacintosh, 2008KalasGraham, 2010LafreniJayawardhana}, the existence of which is 
difficult to explain in the framework of the core accretion mechanism for giant planet formation. Indeed, the timescales for the
accumulation of giant planet atmospheres around solid protoplanetary cores in the core accretion 
models are much longer than the typical age of gaseous disks at distances beyond 10~AU \citep[e.g.,][]{2016StoklDorfi}.

Recently yet another planet formation mechanism has emerged, which combines elements of
both disk fragmentation and core accretion scenarios. In this mechanism, known as tidal downsizing,
gaseous clumps that form in the disk outer regions through gravitational fragmentation migrate inward,
accumulating solid cores in their interior and losing part (or all) of their gaseous atmospheres 
through stellar tidal torques.
The exact outcome of inward migration depends on many parameters, such as the 
migration speed of the clumps, the efficiency of dust settling in the clump interiors, 
the thermodynamics of the clumps, etc., and the final product can be a population of giant 
planets, icy planets, and even terrestrial-like planets in the inner
disk. The key elements of tidal downsizing have been developed
over the past several decades. The possibility for the formation of solid cores inside gaseous 
clumps was  suggested by, e.g.,  \citet{1979Decampli}, \citet{1998Boss} and \citet{2010BoleyHayfield}, and the inward migration,
ejection and survival of gaseous clumps were investigated by \citet{2005VorobyovBasu,2010VorobyovBasu}, \citet{2012ZhuHartmann}, and \citet{2013Vorobyov}.
The refined theoretical foundation for the tidal downsizing
theory was finally put forward in a series of papers by \citet{2010Nayakshinb,2010Nayakshinc,2017Nayakshin}. 

An interesting implication of the tidal downsizing theory is that the inward migration of gaseous 
clumps can result not only in the formation of planets, but also in accretion bursts 
similar in magnitude to FU-Orionis-type eruptions when gaseous clumps are tidally destroyed and accreted
by the protostar. Originally suggested in a series of papers by \citet{2005VorobyovBasu,2010VorobyovBasu,VorobyovBasu2015}, the idea
of FU-Orionis-type eruptions triggered by inward clump migration and tidal destruction
was further investigated semi-analytically by \citet{2012NayakshinLodato} and numerically using 2D thin-disk and 3D numerical (magneto)hydrodynamics simulations by \citet{2011MachidaInutsuka}, \citet{2013VorobyovDeSouza},
\citet{2017MeyerVorobyov}, \citet{2018ZhaoCaselli}, and \citet{2018WhelanRiaz}, showing that that this mechanism can operate in both low- and high-mass
star formation, and also in the primordial stars. 

In this paper, we investigate the process of clump migration in a gravitationally unstable disk using high-resolution numerical hydrodynamics simulations. The use of the thin-disk limit allows us 
to compute the disk evolution during the entire embedded phase of star formation and
achieve the numerical resolution as small as 0.1~AU.
We focus on the properties of one of the gaseous clumps formed in the disk via gravitational 
fragmentation and investigate the details of its inward migration until it is destroyed by 
the action of stellar tidal torques.  We emphasize that we do not use or insert sink particles as proxies
for gaseous clumps, as, e.g., in \citet{2011BaruteauMeru} and \citet{2015Stamatellos}, but instead study
the evolution and dynamics of thermally and rotationally supported, self-gravitating gaseous clumps.
This allows us to study their internal structure, mass loss, and prospects of planetary core formation, as in, e.g.,  \citet{2011ChaNayakshin,2012GalvagniHayfield,2014GalvagniMayer}.

The organization of this paper is as follows. A brief description
of the numerical model is provided in Sect.~\ref{model}. The global disk evolution is considered in Sect.~\ref{sec:global}. The migration of gaseous clumps
is investigated in Sect.~\ref{sec:Dynamics-of-fragment}. The effects of the boundary ans stellar motion are studied in Sect.~\ref{effects}. The main results are summarized in Sect.~\ref{conclude}.

%showing that 
%under certain conditions the infalling clumps can indeed trigger strong accretion bursts
%
%The recent numerical simulation of ... confirmed ... that this process can operate not only in 
%young low-mass stars, but also ...
%the efficiency of tidal truncation of gaseous envelopes around forming protoplanets and 

\section{Model description}
\label{model}

Our numerical model was described in details in \citet{2010VorobyovBasu} and, with some modifications,
in \citet{VorobyovBasu2015}. Below, we briefly review the aspects of the model that are 
most relevant to the present work.
We use numerical hydrodynamics simulations in the thin-disk limit to compute
the formation and long-term evolution of a circumstellar disk.
Our numerical simulations start from the gravitational collapse of
a starless core with a typical radius of $10^{4}$ AU, continues
into the embedded phase of star formation, when a star, disk, and envelope
are formed. The simulations are terminated in the T~Tauri phase, when most
of the envelope has accreted onto the central star plus disk system.
The main physical processes taken into account are  disk heating via stellar and background 
irradiation of the disk surface, shock heating, turbulent heating described via 
the Shakura \& Sunyaev $\alpha$-parameterization, disk self-gravity, and radiative cooling 
from the disk surface. To avoid too small time steps, we set a dynamically inactive sink cell in the center of our computational domain with a radius of $r_{\mathrm{sc}}=15$~AU and introduce
a point-mass protostar in the center of the sink cell when the gas surface density there exceeds a critical
value for the transition from isothermal to adiabatic evolution.
%The sink cell is dynamically inactive; it contributes only to the total gravitational potential 
%and secures a smooth behaviour of the gravity force down to the stellar surface. 
%We also introduce
%a central point-mass object at the center of the sink cell at the
%moment when the gas surface density in the sink cell exceeds a critical
%value for the transition from isothermal to adiabatic evolution. We
%make an assumption, that during the evolution 90\% of the gas that
%crosses the inner boundary of the sink cell is landed onto the central
%star. The other 10\% of the accreted gas is assumed to be carried
%away with protostellar jets.

The equations of mass, momentum, and energy transport in the thin-disk limit are:
\begin{equation}
\frac{\partial\Sigma}{\partial t}=-\bl{\nabla}_{p}\cdot\left(\Sigma \bl{v}_{p}\right),\label{eq:mass}
\end{equation}

\begin{equation}
\frac{\partial}{\partial t}\left(\Sigma \bl{v}_{p}\right)+\left[\bl{\nabla} \cdot \left(\Sigma 
\bl{v}_{p}\otimes \bl{v}_{p}\right)\right]_{p}=-\bl{\nabla}_{p} P+\Sigma \bl{g}_{p}+\left(\bl{\nabla}\cdot
\bl{\Pi}\right)_{p},\label{eq:momentum}
\end{equation}

\begin{equation}
\frac{\partial e}{\partial t}+\bl{\nabla}_{\mathrm{p}}\cdot\left(e v_{p}\right)=-P\left(\bl{\nabla}_{p} \cdot v_{p}\right)-\Lambda+\Gamma+\left(\bl{\nabla} \cdot v\right)_{pp'}: \bl{\Pi}_{pp'},
\label{eq:energy}
\end{equation}
where subscripts $p$ and $p'$ refer to the planar components $(r,\phi)$
in polar coordinates, $\Sigma$ is the mass surface density, $e$
is the internal energy per surface area, $P$ is the vertically integrated
gas pressure calculated via the ideal equation of state as $P=(\gamma-1)e$
with $\gamma=7/5$, $\bl{v}_{p}=v_{\mathrm{r}}\hat{\bl{r}}+v_{\mathrm{\phi}}\hat{\bl{\phi}}$
is the velocity in the disk plane, $\bl{g}_{p}=g_{\mathrm{r}} \hat{\bl{r}} 
+ g_{\mathrm{\phi}} \hat{\bl{\phi}}$
is the gravitational acceleration in the disk plane and 
$\bl{\nabla}_{\mathrm{p}}=\hat{\bl{r}}\partial/\partial r+\hat{\bl{\phi}}r^{-1}\partial/\partial\phi$
is the gradient along the planar coordinates of the disk. 
Viscosity enters the basic equations via the viscous stress
tensor $\bl{\Pi}$ and we calculate the magnitude of kinematic viscosity $\nu$ using the 
$\alpha$-parameterization with a uniform $\alpha=0.01$.

The cooling and heating rates $\Lambda$ and $\Gamma$ take the disk
cooling and heating due to stellar and background irradiation
into account based on the analytical solution of the radiation transfer
equations in the vertical  direction
\citep[see][for detail]{2016DongVorobyov}\footnote{The cooling and heating rates in \citet{2016DongVorobyov}
are written for one side of the disk and need to be multiplied by a factor of 2.}:
\begin{equation}
\Lambda=\frac{8\tau_{\rm P} \sigma T_{\rm mp}^4 }{1+2\tau_{\rm P} + 
{3 \over 2}\tau_{\rm R}\tau_{\rm P}},
\end{equation}
where $T_{\rm mp}={\cal P} \mu / {\cal R} \Sigma$ is the midplane
temperature,  $\mu=2.33$ is the mean molecular weight,  $\cal R$ is the
universal  gas constant, $\sigma$ the Stefan-Boltzmann constant, 
$\tau_{\rm R}$  and $\tau_{\rm P}$ are the  Rosseland and Planck
optical depths to the disk midplane, and  $\kappa_P$ 
and $\kappa_R$ (in cm$^{2}$~g$^{-1}$) are the Planck and Rosseland mean opacities taken from
\citet{2003SemenovHenning}.

The heating function per surface area of the disk is expressed as
\begin{equation}
\Gamma=\frac{8\tau_{\rm P} \sigma T_{\rm irr}^4 }{1+2\tau_{\rm P} + {3 \over 2}\tau_{\rm R}\tau_{\rm
P}},
\end{equation}
where $T_{\rm irr}$ is the irradiation temperature at the disk surface 
determined from the stellar and background black-body irradiation as
\begin{equation}
T_{\rm irr}^4=T_{\rm bg}^4+\frac{F_{\rm irr}(r)}{\sigma},
\label{fluxCS}
\end{equation}
where $F_{\rm
irr}(r)$ is the radiation flux (energy per unit time per unit surface
area)  absorbed by the disk surface at radial distance  $r$ from the
central star. The latter quantity is calculated as 
\begin{equation}
F_{\rm irr}(r)= \frac{L_\ast}{4\pi r^2} \cos{\gamma_{\rm irr}},
\label{fluxF}
\end{equation}
where $\gamma_{\rm irr}$ is the incidence angle of radiation arriving at
the disk surface (with respect to the normal) at radial distance $r$. The
incidence angle is calculated using a flaring disk surface as described
in \citet{2010VorobyovBasu}. The stellar luminosity $L_\ast$ is the sum of the
accretion luminosity  $L_{\rm \ast,accr}=(1-\epsilon) G M_\ast \dot{M}/2
R_\ast$ arising from the gravitational energy of accreted gas and the
photospheric luminosity $L_{\rm \ast,ph}$ due to gravitational
compression and deuterium burning in the stellar interior. The stellar
mass $M_\ast$ and accretion rate onto the star $\dot{M}$ are determined
using the amount of gas passing through the sink cell. 
The properties of
the forming protostar ($L_{\rm \ast,ph}$ and radius $R_\ast$) are
calculated using the stellar evolution tracks provided by the STELLAR code
\citep{2008YorkeBodenheimer}. The fraction of
accretion energy absorbed by the star $\epsilon$ is set to 0.1.

\subsection{Modifications to the original model}
\label{sec:modify}
The main differences with previous studies of \citet{2010VorobyovBasu} and \citet{VorobyovBasu2015}
are 1) an increased numerical resolution, 2) different treatments of the inner boundary condition, 
and 3) stellar motion in response to the gravitational force of the disk.
In this work, we used $1024\times1024$ grid cells, which is a factor of two to four higher 
in each coordinate direction than in our previous works. 
We also increased the radius of the sink cell from 5~AU
to 15~AU, because we are now interested in the dynamics of gaseous clumps (or fragments) at 
distances on the order of several tens to hundreds AU from the star. 
The use of the logarithmically spaced grid in the $r$-direction and equidistant grid 
in the $\phi$-direction allowed us to resolve the disk in the vicinity of the sink
cell with a numerical resolution as small as 0.1 AU and achieve a sub-AU resolution
up to $r=150$~AU. The inner 150~AU are of particular
interest for the current work, because this is where the inward migration
of gaseous fragments formed through disk fragmentation takes place. 
A higher numerical resolution makes it also possible to better resolve the internal structure of 
migrating fragments.

We considered two variants of the inner boundary conditions: the free outflow boundary  
and the free inflow-outflow boundary. In the first case, the matter is allowed to flow out of the active
computational domain, but is not allowed to flow in. This means that zero gradients of 
the gas density, pressure, and radial velocity are applied at the sink -- active disk interface
when $v_r<0$ (the flow is directed towards the sink). When $v_r\ge0$ 
(the flow is directed towards the active disk), the reflecting boundary condition is used.
The azimuthal velocity is extrapolated from the active disk to the sink cell assuming 
a Keplerian rotation.  
The disadvantage of the outflow boundary condition is that it may lead 
to an artificial drop in the gas surface density near the disk
inner edge when wave-like motions are present in the inner disk caused, e.g., by a perturbation from
spiral arms or gaseous clumps orbiting the protostar. 

\begin{figure}
\begin{centering}
\resizebox{\hsize}{!}{\includegraphics{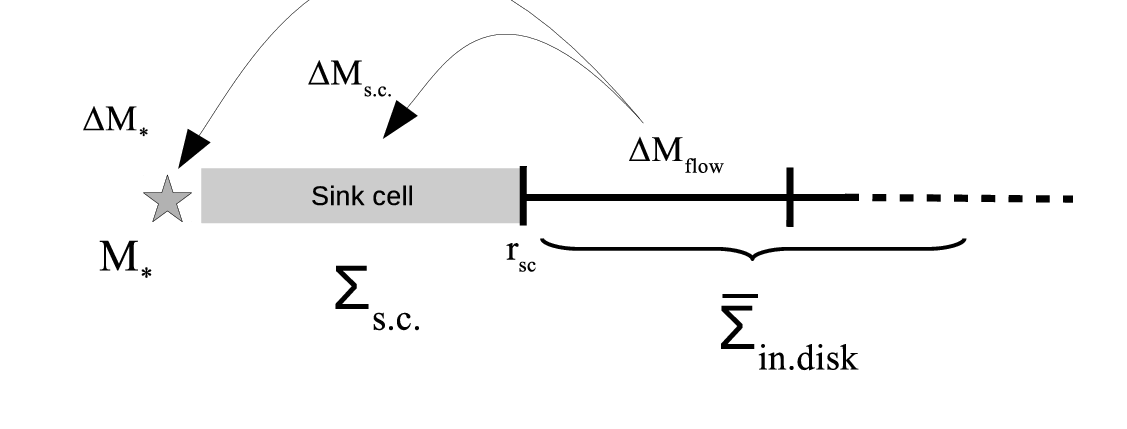}}
\par\end{centering}
\centering{}\protect\protect\protect
\caption{\label{scheme02} 
Schematic illustration of the inner inflow-outflow boundary condition.
The mass of material $\Delta M_{\rm flow}$ that passes through the sink cell
from the active inner disk is further divided into two parts: the mass $\Delta M_\ast$ 
contributing to the  growing central star, and the mass  $\Delta M_{\rm s.c.}$ settling
in the sink cell.
}
\end{figure}

In the inflow-outflow boundary condition, the matter is allowed to
flow freely from the active disk to the sink cell and vice versa.
If the matter flows from the active disk in the sink cell (radial velocity at the interface 
is negative), the mass of material that passes through the sink cell is further redistributed
between the central ptotostar and the sink cell as $\Delta M_{\rm
flow}=\Delta M_\ast + \Delta M_{\rm s.c.}$ (see
Figure~\ref{scheme02}) according to the following algorithm:
\begin{eqnarray}
 \mathrm{if}\,\, \Sigma_{\rm s.c.}^n < \overline{\Sigma}_{\rm in.disk}^n\,\, \mathrm{then} \nonumber\\
 \Sigma_{\rm s.c}^{n+1}&=&\Sigma_{\rm s.c.}^n+\Delta M_{\rm s.c.}/S_{\rm s.c.} \nonumber\\
 M_\ast^{n+1}&=&M_\ast^n+\Delta M_\ast \nonumber \\
 \mathrm{if}\,\, \Sigma_{\rm s.c.}^n \ge \overline{\Sigma}_{\rm in.disk}^n\,\, \mathrm{then} \nonumber\\
 \Sigma_{\rm s.c.}^{n+1}&=& \Sigma_{\rm s.c.}^n \nonumber\\
 M_\ast^{n+1}&=& M_\ast^n + \Delta M_{\rm flow}. \nonumber
\end{eqnarray}
Here, $\Sigma_{\rm s.c.}$ is the surface density of gas in the
sink cell,  $\overline\Sigma_{\rm in.disk}$ the averaged surface density
of gas in the inner active disk (the averaging is usually done  over
several AU immediately adjacent to the sink cell, the exact value is
determined by numerical experiments), and $S_{\rm s.c.}$ the surface area 
of the sink cell. The exact partition between $\Delta M_\ast$ and $\Delta
M_{\rm s.c.}$ is usually set to 95\%:5\%.  If the matter flows from the sink cell to 
the active disk (radial velocity at the interface is positive), 
then we update only the  surface density in the sink cell as
$\Sigma_{\rm s.c.}^{n+1}=\Sigma_{\rm s.c.}^n - \Delta M_{\rm flow}/S_{\rm
s.c.}$ and do not change the mass of the central star.

The calculated surface densities in the sink cell $\Sigma_{\rm
s.c.}^{n+1}$  are used at the next time step as boundary values for the surface
density.  The radial velocity and pressure in the sink cell are determined
from the zero gradient condition, while the azimuthal velocity is extrapolated 
from the active disk to the sink cell assuming a Keplerian rotation.
These  inflow-outflow boundary conditions enable a smooth transition of the surface density
between the inner active disk and the sink cell, preventing (or greatly
reducing) the formation of an artificial drop in the surface  density
near the inner boundary. We ensure that our
boundary conditions conserve the total mass budget  in the system.
Finally, we note that the outer boundary condition is set to a standard 
free outflow, allowing for material to flow out of the computational
domain, but not allowing  any material to flow in.

To explore the effects of stellar motion in response to the gravity force of 
the disk (including fragments), a term $-\Sigma \bl{g}_\ast$ was added
to the right-hand-side of Equation~(\ref{eq:momentum}), so that this equation now describes 
the transport of momentum in
the non-inertial frame of reference moving with the star. The acceleration of the star's frame of reference $\bl{g}_\ast$ can be expressed as
\begin{equation}
\bl{g}_\ast = G \int { dm(\bl{r}^\prime) \over r^{\prime 3} } \bl{r}^\prime,
\label{starAccel}
\end{equation} 
where $dm$ is the mass in a grid cell with position vector $\bl{r^\prime}$.
In practice, we find $\bl{g}_\ast$ by
first calculating its Cartesian components $g_{\ast,x}$ and $g_{\ast,y}$ as
\begin{eqnarray}
g_{\ast,x} &=&  \sum_{j,k} {F}_{j,k} \cos(\phi_{k}), \\
g_{\ast,y} &=&  \sum_{j,k} {F}_{j,k} \sin(\phi_{k}),
\end{eqnarray}
where $\phi_{k}$ is the azimuthal angle of the grid cell $(j,k)$, 
the summation is performed over all grid zones  and the force (per unit stellar mass) 
acting from the grid cell $(j,k)$ onto the star can be expressed in the following form
\begin{equation}
{F}_{j,k}= G {m_{j,k} \over r_j^2 },
\label{accel}
\end{equation}
where $m_{j,k}$ is the mass in the grid cell $(j,k)$ and $r_j$ is the radial distance to the grid cell
$(j,k)$.

Once the Cartesian components $g_{\ast,x}$ and $g_{\ast,y}$ of the stellar acceleration are known 
in every grid cell, the corresponding polar-grid components can be found using the 
standard coordinate transformation formula
\begin{eqnarray}
g_{\ast,r}&=& g_{\ast,x}\cos(\phi) + g_{\ast,y} \sin(\phi),  \\
g_{\ast,\phi}&=& - g_{\ast,x}\sin(\phi) + g_{\ast,y} \cos(\phi),
\end{eqnarray}  
where $\phi$ is the azimuthal coordinate of a given grid cell.
More detail can be found in \citet{2017RegalyVorobyov}.

\begin{figure*}
\begin{centering}
\includegraphics[width=2\columnwidth]{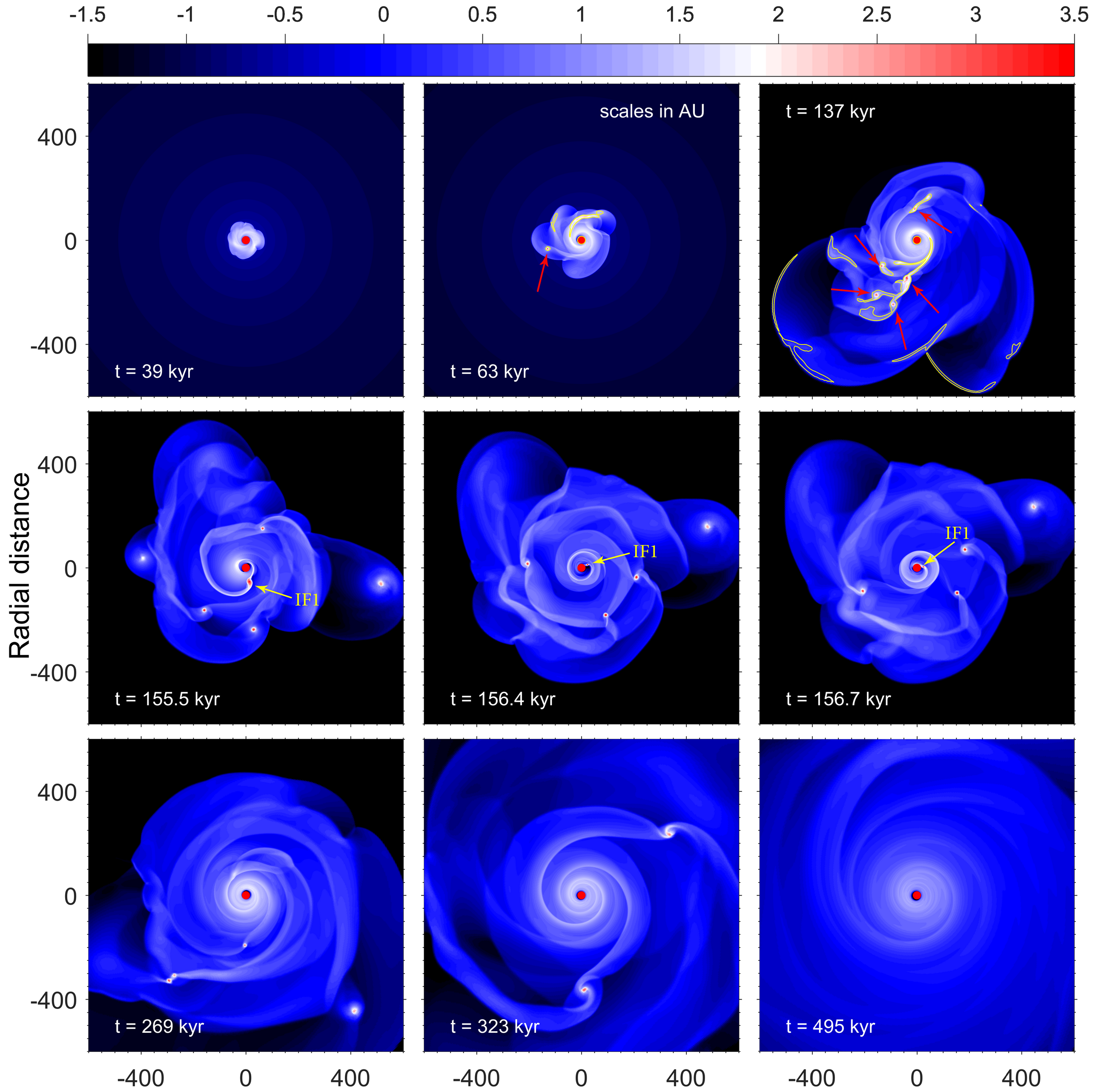}
\par\end{centering}
\caption{\label{fig:2}Gas surface density maps ($\mathrm{g\,cm^{-2}}$,
log units) in the fiducial model shown at nine times since the formation
of the central protostar. Only the inner $600 \times 600$ AU$^{2}$ box
is shown, the total computational region extends  to $14000$~AU in the 
$r$-direction. The contour lines in the top left panel outline the regions 
where the Toomre $Q$-parameter is less than unity and the red arrows point to the identified
fragments in the disk. Fragment IF1 is marked by the yellow arrows in the middle row of panels; 
its dynamics is considered in detail
in Sect.~\ref{sec:Dynamics-of-fragment}.}
\end{figure*}

\subsection{Initial conditions}

The initial radial profile of the gas surface density $\Sigma$ and
angular velocity $\Omega$ of the pre-stellar core has the
following form: 
\begin{equation}
\Sigma=\frac{r_{0}\Sigma_{0}}{\sqrt{r^{2}+r_{0}^{2}}}\label{eq:sigma}
\end{equation}
\begin{equation}
\Omega=2\Omega_{0}\left(\frac{r_{0}}{r}\right)^{2}\left[\sqrt{1+\left(\frac{r}{r_{0}}\right)^{2}}-1\right]\label{eq:omega}
\end{equation}
where $\Sigma_{0}$ and $\Omega_{0}$ are the angular velocity and
gas surface density at the center of the core, $r_{0}=\sqrt{A}c_{\mathrm{s}}^{2}/\pi G\Sigma_{0}$
is the radius of the central plateau, where $c_{\mathrm{s}}$ is the
initial isothermal sound speed in the core. This radial profile
is typical of pre-stellar cores formed as a result of the slow expulsion
of magnetic field due to ambipolar diffusion, with the angular momentum
remaining constant during axially-symmetric core compression (Basu
1997). The value of the positive density perturbation \textit{A} is
set to 1.2, making the core unstable to collapse. The initial gas temperature in collapsing cores is $T_{\mathrm{init}}=10\,\mathrm{K}$.
We consider a numerical model with $\Omega_{0}=1.2$~km~s$^{-1}$~pc$^{-1}$, 
$\Sigma_{0}=5.2\times10^{-2}\,\mathrm{g\,cm^{-2}}$, and $r_{0}=2400\,\mathrm{AU}$.
The resulting core mass $M_{\rm core}=1.1\,M_{\odot}$ and
the ratio of rotational to gravitational energy $\beta=6.1\times10^{-3}$.

\section{ Global disk evolution \label{subsec:Disk-evolution}}
\label{sec:global}
In this section, we analyze the global evolutionary trends that are typical of our model disk, 
while in Sect.~\ref{sec:Dynamics-of-fragment} we focus on a shorter time period and analyze 
the inward 
migration  of one of the gaseous clumps formed in the disk through gravitational
fragmentation. The fiducial model considered in this section and 
Sect.~\ref{sec:Dynamics-of-fragment} is characterized by the free outflow boundary condition 
and the motionless central star. The effects of stellar motion and free inflow-outflow boundary condition
will be considered in Sect.~\ref{effects}.
 
When disk fragmentation takes place, we distinguish
the newly formed fragments from the rest of the disk (e.g., from the spiral arms) using
the fragment-tracking method (first introduced in \citet{2013Vorobyov}) that searches 
for the disk  regions satisfying the following criteria. First, we identify the local 
surface density peaks in the disk and stipulate that they represent the centers of the fragments 
if the gas surface density in these peaks is at least
a factor of 10 higher than the azimuthally averaged gas surface density at the same radial distance
from the star. The exact factor was found using experiments and visual checks.
After the center of the fragment with coordinates ($r_{\mathrm{c}},\varphi_{\mathrm{c}}$) 
has been identified, we determine the neighboring mesh cells belonging
to the fragment by imposing the following two conditions on the gas
pressure $P$ and gravitational potential $\Phi$
\begin{equation}
\frac{\partial P}{\partial r'}+\frac{1}{r'}\frac{\partial P}{\partial\varphi'}<0,
\label{eq:Pgrad}
\end{equation}
\begin{equation}
\frac{\partial\Phi}{\partial r'}+\frac{1}{r'}\frac{\partial\Phi}{\partial\varphi'}>0,
\label{eq:PhiGrad}
\end{equation}
where $r'=r-r_{\mathrm{c}}$ and $\varphi'=\varphi-\varphi_{\mathrm{c}}$.
Equation (\ref{eq:Pgrad}) requires that the fragment is pressure-supported,
with a negative pressure gradient with respect to the center of the
fragment. Equation (\ref{eq:PhiGrad}) requires that the fragment is kept together by gravity, with 
the potential well being deepest at the center of the fragment. If these conditions fail 
at the center of the fragment, then the fragment is rejected,
meaning that we falsely took a local density perturbation for a pressure-supported, gravity-bound fragment.
If these conditions are fulfilled at the center of the fragment, we continue marching 
from the center and checking the neighbouring cells until any of 
the conditions is violated. The grid cells that fulfill both conditions constitute the fragment.   
We checked the validity of this algorithm by random visual checks and found it to be in most cases 
satisfactory.

\begin{figure*}
\begin{centering}
\includegraphics[width=2\columnwidth]{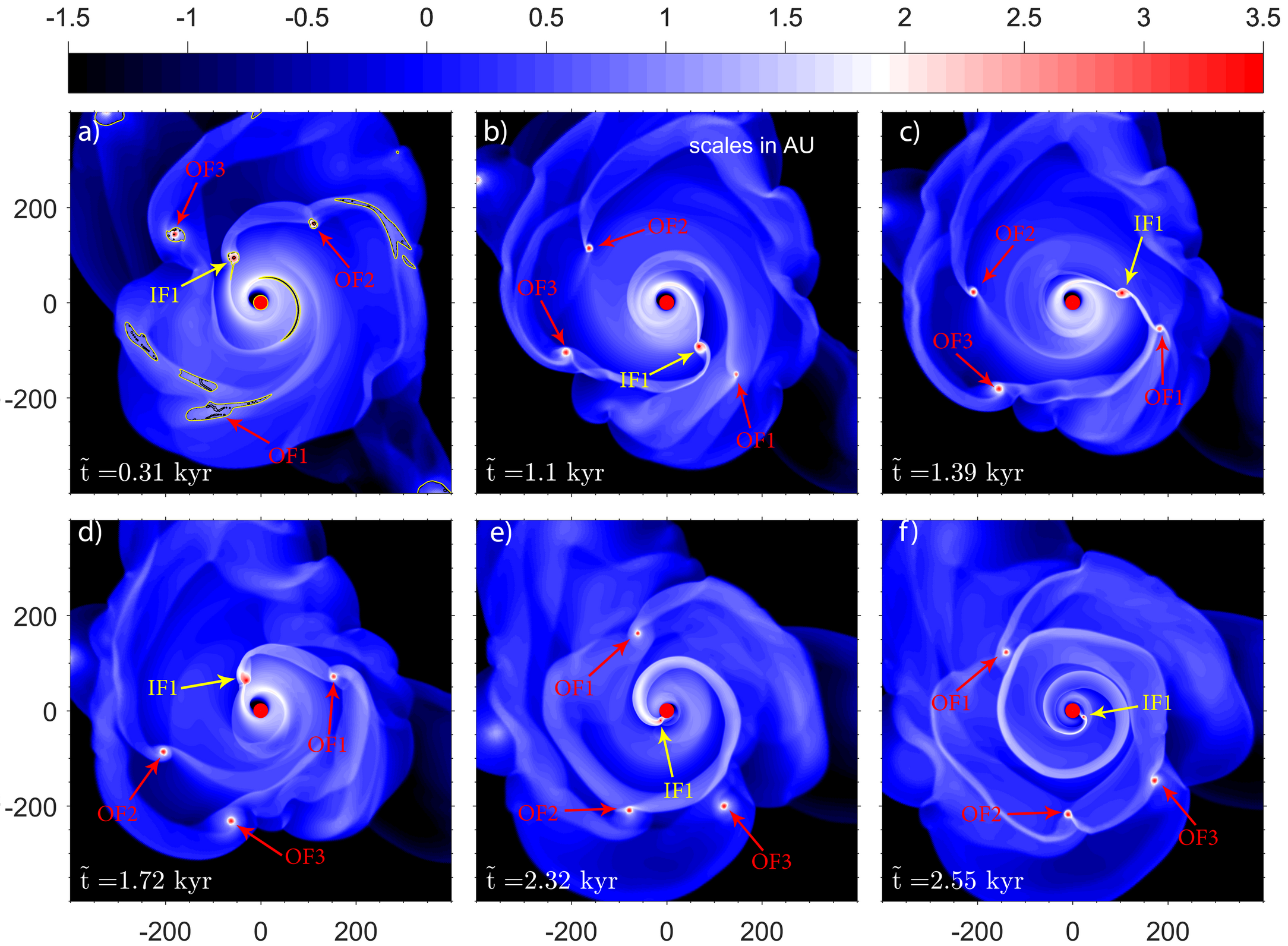}
\par\end{centering}
\caption{\label{fig:3}Gas surface density distributions in the fiducial model 
at several time instances focusing on the inward migration of IF1. The time 
is counted from $t_{0}=153.5$~kyr. The position of IF1 is
marked with the yellow arrows, while the positions of OF1, OF2, and
OF3 are marked with the red arrows. The yellow contour lines outline disk regions with the Toomre $Q$-parameter smaller than unity, while the black contour lines show the disk regions satisfying both the Toomre and Gammie criteria for disk fragmentation.  The disk rotates counter-clockwise.}
\end{figure*}

Figure~\ref{fig:2} presents
the gas surface density images in the inner $600\times600$~AU$^2$ box at nine  evolutionary times,
starting  soon after the disk formation and ending after 0.5~Myr of disk evolution
when most of the initial core mass has accreted on the star plus disk system. We therefore capture the
entire embedded phase of disk evolution.
All times in the current paper are calculated from the instance of central star formation,
if not stated otherwise, and the disk forms at $t=29$~kyr.
The fast increase in the disk size is evident in the top row. The first fragment forms 
at $t\approx60$~kyr at a distance of $\approx 100$~AU, in agreement with other
studies of gravitationally unstable disks showing that disk fragmentation is most 
likely to occur in the disk outer regions, $r\ga 50$~AU \citep[e.g.][]{2009StamatellosWhitworth,2009Boley}.
In the subsequent evolution, more fragments emerge in the disk, as shown by the red arrows, 
but their number does not grow steadily with time due to the presence of several mechanisms 
that lead to the loss of fragments. These mechanisms include merging of 
two fragments due to close encounters, ejection of fragments from the
disk due to multi-body gravitational interaction \citep[e.g.,][]{2012BasuVorobyov,2016Vorobyov},
and inward migration of fragments followed by their tidal destruction
and accretion on the central star 
\citep{2010VorobyovBasu,2011ChaNayakshin,2012ZhuHartmann,VorobyovBasu2015}. 
The middle row of panels in Figure~\ref{fig:2} is dedicated to a particular fragment,
referred as IF1, which properties and inward migration is studied in detail in 
Sect.~\ref{sec:Dynamics-of-fragment}.

To check if the Toomre criterion for disk fragmentation is fulfilled in our model, we 
calculate the Toomre $Q$-parameter and search for the disk regions where $Q<1.0$ 
\citep{1964Toomre}. The $Q$-parameter for the case of a Keplerian disk is defined as
\begin{equation}
Q={c_{\mathrm{s}}\Omega \over \pi G\Sigma},
\end{equation}
where $c_{\mathrm{s}}$ is the local sound speed, $\Omega$
and $\Sigma$ are the local angular velocity and gas surface
density in the disk, respectively. 
The yellow contour lines in the top row of Figure~\ref{fig:2} outline the 
regions of the disk where the Toomre parameter is less than unity and the
red arrows point to the identified fragments. Clearly, all fragments are
surrounded by compact disk regions with $Q<1.0$, fulfilling the Toomre fragmentation criterion.
At the same time, there exist also disk regions that are characterized by $Q<1$, but containing 
no fragments. These regions may form fragments later. 
We note that in the early evolution  at $t<140$~kyr all fragments are located at 
radial distances $r>100$~AU, where disk fragmentation is most likely.
In the later evolution, however, some fragments are seen at much smaller
radial distances, on the order of several tens of AU. Although disk fragmentation 
at these distances is possible if induced by fragments already present is the disk \citep{2015Meru},
in our case this is the result of inward migration and scattering, which we discuss in more detail 
in Sect.~\ref{sec:Dynamics-of-fragment}.

Figure~\ref{fig:2} demonstrates that no fragments are present in the disk at the end of numerical simulations ($t=0.495$~Myr), meaning that none of the previously formed fragments have survived through the embedded
phase of disk evolution. This indicates that although disk fragmentation can be common
in the embedded phase, the survivability of the fragments 
is low \citep[see, e.g.,][]{2013Vorobyov} and most, if not all, fragments are destroyed by the
end of the embedded phase.
There is, however, one caveat to this conclusion. When considering the evolution of fragments
we did not take into account their possible contraction to planetary-sized objects. This occurs 
if the gas temperature exceeds 2000~K, molecular hydrogen dissociates, and
the interiors of the fragment experience fast contraction due to the loss of pressure support
\citep[see, e.g.,][]{2000MasunagaInutsuka}. Resolving this process requires a much higher numerical resolution than is affordable in our core collapse and disk formation simulations. Alternatively,
one can introduce sink particles as proxies for the fragments,
a practice often adopted in studies of disk fragmentation 
\citep[see, e.g.,][]{2010FederrathBanerjee}. 
In this case, however, one cannot resolve the internal
structure of the fragments. In Sect.~\ref{sec:Dynamics-of-fragment}, we analyze 
the consequences of our adopted approximation and demonstrate that some fragments may give birth
to dense planetary-sized objects before being finally destroyed by the action of stellar tidal torques.

%To summarize, the disk evolution in our higher resolution simulations demonstrates
%features (such as fragmentation) that were also typical for our lower resolution runs 
%\citep[e.g.,][]{2010ApJ...719.1896V,VorobyovBasu2015}. 
%However, our higher resolution simulations allowed us to uncover important details of 
%the fragment migration process, which are discussed in detail in the next section.
%In the next section, we analyze in detail the structure and migration of fragment IF1.

\section{Dynamics of fragment IF1} 
\label{sec:Dynamics-of-fragment}

The simultaneous presence of several fragments in the disk may greatly
complicate their dynamics and evolution. In this section, the dynamics
of four fragments, denoted further in the text as IF1 (inner fragment
1), OF1, OF2, and OF3 (outer fragments 1, 2, and 3, respectively), 
are studied in detail to show how the interaction between different 
fragments occurs in the disk and what may be the consequences of this interaction. 
We particularly focus on the structure and dynamics of IF1, as it shows 
the most 
interesting behavior when approaching the central protostar.

Figure~\ref{fig:3} presents the disk surface density distributions at six
consecutive times starting from the time instance when OF1 forms in the disk and 
ending when IF1 halts its inward migration and settles on a quasi-stable orbit.
In the following text, the time is counted from $t_0=153.5$~kyr and is referred
as $\tilde{t}=t-t_0$.
%the gravitational interaction between IF1 and the other three fragments OF1, OF2 and OF3 ensues. 
%The disk rotates counterclockwise. 
The yellow arrows in Figure \ref{fig:3} mark IF1, while the red arrows
mark OF1, OF2 and OF3. Panel~a) demonstrates that OF1 starts forming at $\tilde{t}=0.31$~kyr 
from a local density enhancement that fulfills the Toomre fragmentation criterion, 
as can be seen from the yellow contour lines outlining the disk regions with $Q<1$. 
OF1 is fully formed already at $\tilde{t}=1.1$~kyr and its
mass at this time instance is 4.5 Jovian masses. We also checked if
the regions with $Q<1$ satisfies the Gammie criterion \citep{2001Gammie} stating
that the local cooling time $t\mathrm{_{cool}}$ should be shorter
than the fastest growth time of gravitational instability 
$t_{\mathrm{grav}}=2\pi/(\Omega\sqrt{1-Q^{2}})$.
The disk regions that satisfy both the Gammie and Toomre criterions
are outlined with the black contour lines in panel a) Figure \ref{fig:3}.
Clearly, these regions are more compact than the regions outlined by the $Q<1$ criterion only,
meaning that these two criteria, when applied together, are mores stringent than
the Toomre criterion alone. Nevertheless, the regions where both criteria are fulfilled
are present in several disk locations and the formation of OF1 fulfills both criteria.

\begin{figure}
\begin{centering}
\includegraphics[width=1\columnwidth]{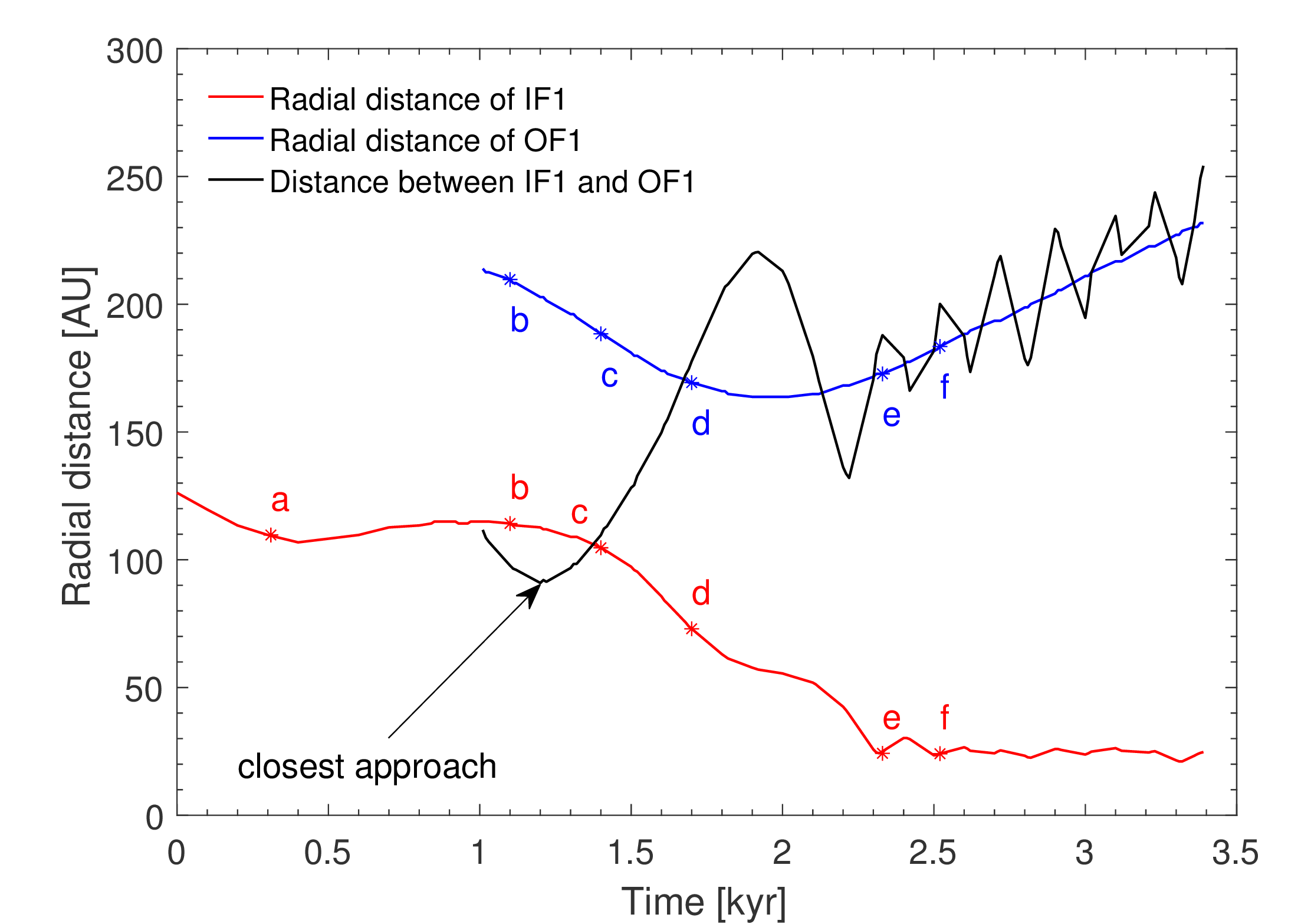}
\par\end{centering}
\caption{\label{fig:4} Radial distances of IF1 and OF1 from the protostar
(the red and blue lines, respectively) as a function of time. 
The black line shows
the relative distance between IF1 and OF1. The letters indicate
the time instances corresponding to panels b), c), d), e), and f) 
in Figure \ref{fig:3}. 
%Time moment $t=t_{0}+0.31$
%kyr of the panel (a) from Figure \ref{fig:3} is not shown in Figure
%\ref{fig:4} due to the fact that the fragment OF1 has not yet formed.
}
\end{figure}

To check if the mass of OF1 is in agreement with theoretical
expectations, we utilized equation~(3) from \citet{2013Vorobyov} expressing
the Jeans length $R_{\mathrm{J}}$ in a thin disk as
\begin{equation}
R_{\mathrm{J}}=\frac{\left\langle v^{2}\right\rangle }{\pi G\Sigma},
\end{equation}
where $\left\langle v^{2}\right\rangle =2RT_{\rm mp}/\mu$ is the two-dimensional 
velocity dispersion.   
To find $R_{\rm J}$, we first alculated the mean values 
of the gas surface density and midplane temperature ($\overline \Sigma$ and $\overline{T}_{\rm mp}$)
inside the region with $Q<1$ centered on the forming OF1. 
The resulting mass of OF1, calculated as
$M_{\mathrm{fr}}=\pi R_{\mathrm{J}}^{2}\overline{\Sigma}$,  is equal to 
12.7~$M_{\rm Jup }$, which is a factor of three higher than what was found
using the fragment-tracking method. To resolve this discrepancy, we 
note that the region that finally collapses to form OF1 may be smaller than
that outlined by the $Q<1$ condition.  A good agreement is found if we consider the disk region around the
forming fragment that fulfills a more stringent $Q<0.5$ condition on disk fragmentation. A more stringent Toomre condition for fragmentation 
was also reported in the recent work of \citet{2016TakahashiTsukamoto}.

%where $\Sigma\mathrm{_{mean}}$ is the mean surface density of the
%gas inside the yellow contour lines. The mass of fully formed Fragment
%OF1 is about 4.5 Jovian masses and is in a good agreement with the
%theoretically predicted mass of the fragment OF1 differing with less
%than a factor of 3. 

Panels b) and c) in Figure \ref{fig:3} show
the time instances when OF1 has fully formed and it starts interacting gravitationally 
with IF1.  A spiral arc connecting both fragments develops during the closest approach between the two fragments and the angular momentum is redistributed so that IF1 starts spiralling toward the central protostar and OF1 is pushed to a
higher orbit.

\begin{figure*}
\begin{centering}
\includegraphics[width=2\columnwidth]{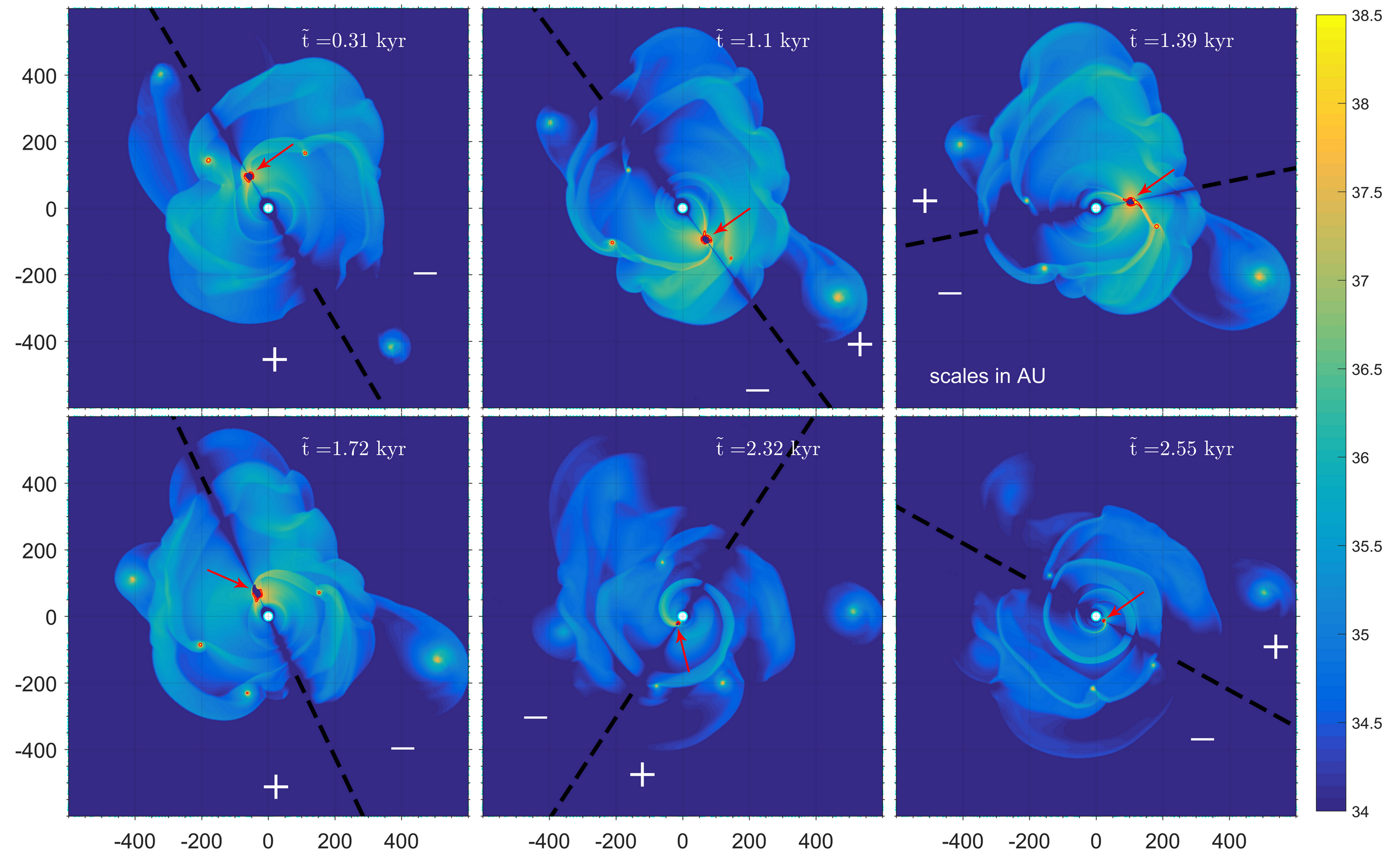}
\par\end{centering}
\caption{\label{fig:6} Spatial maps of the absolute torque exerted
on IF1 by the rest of the disk. The times are similar to those shown in Fig.~\ref{fig:3}. 
The disk rotates counterclockwise. The red arrows point to IF1. 
Parts of the disk characterized by the 
positive and negative torque are marked with the ``+'' and ``-'' signs, respectively, and the black dashed line separates these parts. The color bar shows the torque 
values in log dyne cm. The red contour line highlights the disk regions where the torque exceeds $10^{38}$ dyne cm. }
\end{figure*}

The interaction between IF1 and OF1 is illustrated in Figure \ref{fig:4} showing 
the radial distances of OF1 and IF1 from the protostar as a function of time. 
We also show the relative distance between OF1 and IF1, which changes with time because 
of migration and different angular velocities of the fragments. The time instance of the closest 
approach between IF1 and OF1 is marked with the arrow in Figure \ref{fig:4}. 
%Blue and
%red solid curves in Figure \ref{fig:4} show the radial distances
%of fragments OF1 and IF1, respectively, from the central protostar.
%Markers on the red and blue curves represent the time moments at which
%respective panels in Figure \ref{fig:3} are shown. 
%Close interaction of fragments IF1 and OF1 starts almost immediately
%after the formation of fragment OF1, as soon as the later forms in
%the vicinity of fragment IF1. 
Before the formation of OF1 ($\tilde{t}<1.0$~kyr), IF1 orbited 
the protostar at a distance of 110--120~AU.
Soon after the closest approach between OF1 and IF1, 
the radial distance of the latter starts rapidly
decreasing, indicating that it lost angular momentum during this event,
while the radial distance of OF1 starts increasing, implying that it gained
angular momentum.
Interestingly, IF1 halts its inward migration at $\tilde{t}\approx 2.3$~kyr and starts orbiting the protostar 
at a radial distance of $\approx 25$~AU.
%IF1 stays at this orbit for more than 1000 years before
%it loses the bulk of its initial mass through tidal torques and finally disperses.
The reasons that cause IF1 to halt its inward migration 
are discussed in detail in Sect.~\ref{subsec:Halt-of-migration}.

The close encounter of IF1 with OF1 played the role of a trigger,
which initiated a rapid inward migration of IF1.
However, other fragments in the disk, such as OF2 and OF3, or other structures 
in the disk, such as spiral arms and local density enhancements, probably also played 
an important role in the inward migration of IF1. To determine their
influence on the dynamics of IF1, we calculated
the gravitational torques $\tau_i$ exerted on IF1 by the $i$-th grid cell of the disk
as
\begin{equation}
\tau_{i}=|\bl{r}_{\mathrm{f}}| \, |\bl{F}_{i}| \, \sin\gamma_{i},
\end{equation}
where 
$|\bl{r}_{\mathrm{f}}|$ is the distance between the center of the fragment
and the protostar (i.e., the lever arm),  $|\bl{F}_{i}|$ is the gravitational
force acting on the fragment from the $i$-th cell, and 
$\gamma_{i}$ is the angle between the lever arm and the force.
When calculating $\tau_{i}$, we excluded the grid cells 
belonging to IF1 itself. The gravity force is calculated as
\begin{equation}
|\bl{F}_{i}|={G m_{i} M_{\rm f} \over R_i^{2} }
\end{equation}
where $M_{\mathrm{f}}$ is the
mass of the fragment, $m_{i}$ is the mass inside the $i$-th
grid cell, and $R_i$ is the distance between the $i$-th
cell and the center of IF1.

The total torque  $\bl{\tau}_{\rm tot} = \sum_i \bl{\tau}_{i}$ acting on IF1 can be related 
to its angular momentum as
\begin{equation}
{d \bl{L} \over dt } = \bl{\tau}_{\rm tot}
\end{equation}
and can therefore give us an insight into the radial migration of the fragment
-- the negative $\tau_{\rm tot}$ would imply that the fragment is losing its angular momentum and approaching
the star, while the positive $\tau_{\rm tot}$ would imply the opposite.

Figure~\ref{fig:6} presents the spatial maps of individual torques $|\bl{\tau}_i|$ 
(by the absolute value) exerted  on IF1 by each grid cell of the disk. The 
position of IF1 is marked with the red arrow and the times
are similar to those in Figure \ref{fig:3}. 
The fragment itself is highlighted by the black color to emphasize that the grid cells constituting the fragment do not contribute to the calculation of $\bl{\tau}_{\rm tot}$.
Clearly,
$\tau_i$ is equal to zero for the grid cells located at the line passing 
through the star and the center of IF1 (because $\sin\gamma_{i}=0$). This line
divides the disk into two halves, each one characterized by a distinct sign of the torque, 
positive or negative, that they exert on IF1.
%Moreover, the individual torques $\tau_i$ have opposite signs on each have of the disk 
To distinguish between parts of the disk with positive and negative torques, we used the ``+'' 
and ``--'' signs, respectively. We note that IF1 orbits the star in the
counterclockwise direction. Therefore, torques from part of the disk 
that is located ahead of the IF1 direction of rotation are positive.

\begin{figure}
\begin{centering}
\includegraphics[width=1\columnwidth]{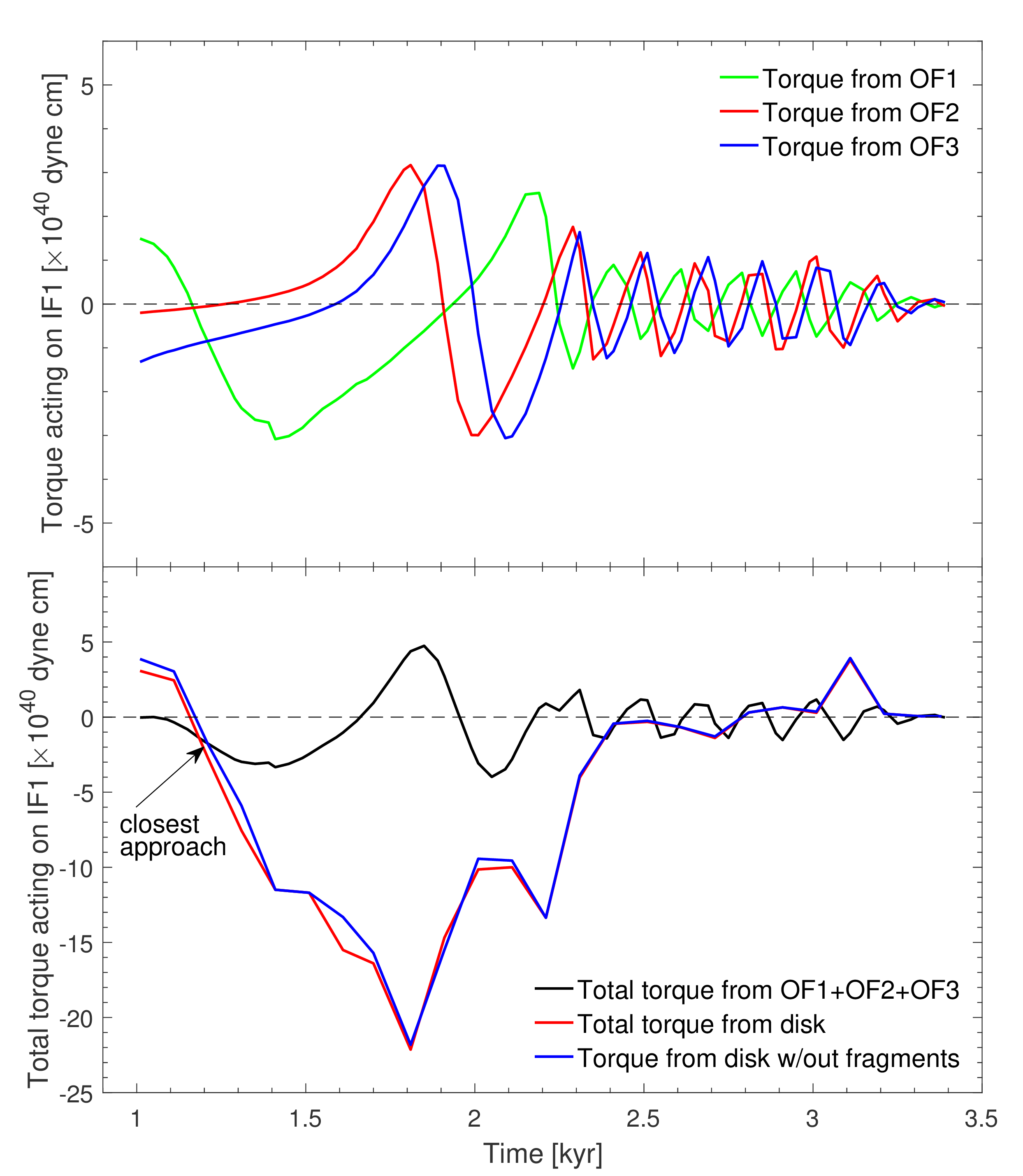}
\par\end{centering}
\caption{\label{fig:7}{\bf Top panel}: Torque exerted on IF1 from OF1 (the green line),
OF2 (the red line), and OF3 (the blue line) vs. time. {\bf Bottom panel}: Torque
exerted on IF1 by OF1, OF2, and OF3 taken together (the black
line), by the entire disk (the red line), and by the entire disk, but excluding 
the torque from the fragments (the blue line) vs. time. Note the difference in scales.}
\end{figure}

A comparison of Figs.~\ref{fig:6} and \ref{fig:3} indicates that the disk regions with 
the highest gas surface density, such as fragments and spiral arms, exert strongest torques on 
IF1. Moreover, the torque exerted by OF1 is evidently not the strongest. For example, the 
arc connecting OF1 and IF1 at $\tilde{t}=1.39$~kyr exerts a comparable torque on IF1. 
This supports our previous
conjecture that the close approach of OF1 with IF1 acts only as a trigger, which initiates
the fast inward migration of IF1, and the input from the entire disk needs to be considered 
when analyzing the migration of fragments in a self-gravitating disk.
We also note that the torques acting on IF1 appear to decrease by absolute value at later times 
as IF1 approaches the star. 

%torque acting on Fragment IF1 from the regions of the disk with high
%surface density, such as spiral arms or fragments, is higher by an
%order of magnitude than from the rest of the disk. Top right panel
%in Figure \ref{fig:6} shows how strong is the negative torque from
%Fragment OF1 and the ``bridge'' between Fragments IF1 and OF1. After
%the interaction with OF1, Fragment IF1 starts to loose its mass and
%forms a spiral track from the matter lost. The torques from the spiral
%tracks can be clearly seen in the bottom panels of Figure \ref{fig:6}.

In Figure~\ref{fig:7} we compare the gravitational torques exerted on
IF1 by OF1, OF2 and OF3, and also by the entire disk.
More specifically, the top panel shows the torques exerted by each of the three
fragments separately, while the bottom panel shows the torque exerted by  
the three fragments taken together, by the entire disk, and by the entire disk, 
but excluding the three fragments. 
Clearly, the torques from each of the three fragments (OF1, OF2, and OF3) 
are comparable by absolute value, but appear to be "out of phase", 
partly cancelling each other. As a result, the cumulative torque from the three 
fragments (the black line in the bottom 
panel) has a sinusoidal shape with an amplitude that gradually declines with time. 
We note that after the closest approach between IF1 and OF1 (at $\tilde{t}\approx 1.2$~kyr), 
the disk torque becomes both negative and 
much higher by absolute value than the cumulative torque from the three fragments.
%The input of the fragments in the total torque from the entire disk is rather small,
%as the comparison of blue and red circles demonstrates.
The subsequent inward migration of IF1 is therefore sustained by the disk 
rather than by the fragments. The main input to the disk torque may be from the spiral arc
that connects OF1 with IF1 during their closest approach ($\tilde{t}=1.39$~kyr) and later 
transforms in a spiral tail, which is visible in the wake of IF1 at $\tilde{t}=1.72$~kyr and 
$\tilde{t}=2.32$~kyr.

\begin{figure*}
\begin{centering}
\includegraphics[width=2\columnwidth]{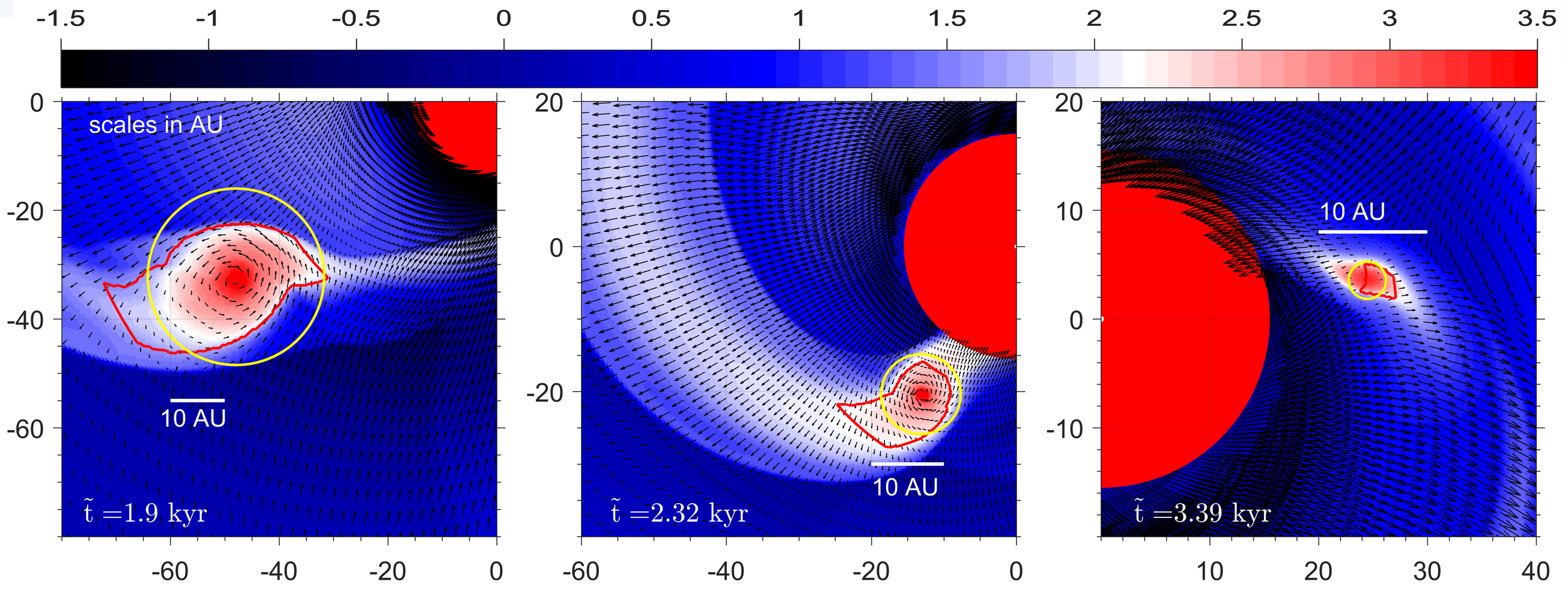}
\par\end{centering}
\caption{\label{fig:8}Zoomed-in view on IF1 during its inward migration
towards the protostar. The black arrows
show the gas velocity field superimposed on the gas surface density distribution.
The shrinking spatial scales are used to better resolve the fragment.
The yellow circles mark the Hill radius of the fragment and the red curves 
outline the fragment shape as determined by the fragment-tracking method.
The color bar shows the gas surface density in log~g\,cm$^{-2}$.}
\end{figure*}

%During the inward migration fragments become elongated because of
%acting on them gravitational and tidal torques, forming shapes similar
%to the spiral arms. 
Figure \ref{fig:8} shows the gas surface density distribution in and around IF1 
during its inward migration towards the protostar. Three different time instances are shown 
and every panel has progressively smaller spacial scales to better resolve the fragment. 
The black arrows show the gas velocity field in the frame of reference of the center of IF1. 
Clearly, IF1 rotates in the same direction as the disk.
The arrows for every 30-th grid cell in each coordinate direction were only shown to
avoid cluttering. The shape of IF1, found with the fragment tracking algorithm 
of Sect.~\ref{sec:global}, is outlined with the red curves.
The yellow circles outline the Hill radius for IF1 calculated as 
\begin{equation}
R_{\mathrm{H}}=r_{\mathrm{f}}\left(\frac{M_{\mathrm{f}}}{3(M_{*}+M_{\mathrm{f}})}\right)^{1/3},
\end{equation}
where $M_{\mathrm{f}}$ is the mass of the fragment confined within the red curve. 

\begin{figure}
\begin{centering}
\includegraphics[width=1\columnwidth]{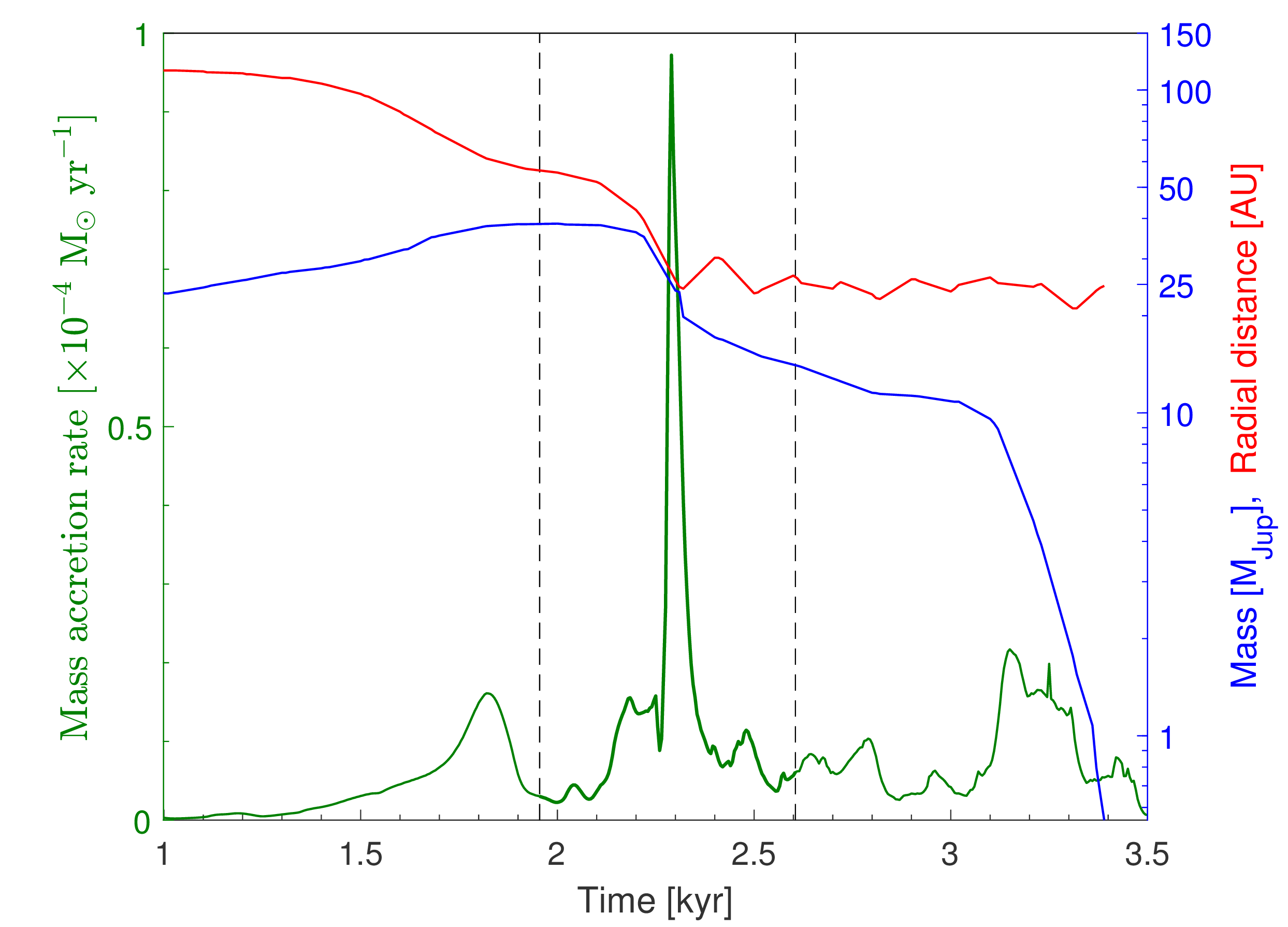}
\par\end{centering}
\caption{\label{fig:10} Mass accretion rate through the sink cell (the green line),
radial distance of IF1 from the star (the red line), and mass of IF1 (the blue line) 
as a function of time. }
\end{figure}

As IF1  approaches the star, its size shrinks because of the shrinking Hill radius 
and its shape becomes distorted  through the action of tidal torques. 
Simultaneously, part of its material starts streaming away along the tidal arms.
As a result, the fragment loses a large fraction of its initial mass
through the leading and trailing tidal arms. 
Finally, the rotational motion inside IF1 almost disappears ($\tilde{t}$=3.39~kyr), 
turning into a sheer outflow. At this stage, IF1 ceases to be gravitationally supported 
and tidal torques tear apart the fragment.

The tidal destruction of IF1 is accompanied by an accretion burst with 
a magnitude typical of the FU-Orionis-type eruptions \citep{2014AudardAbraham}.
Figure \ref{fig:10} shows  the mass accretion rate $\dot{M}$
through the sink cell, the mass of IF1, and the radial distance of IF1 from 
the protostar as a function of time.
As the fragment migrates inward, it first accumulates mass (reaching a peak value of $\approx 38~M_{\rm Jup}$),
but then rapidly loses a large fraction of its mass when approaching a distance of 25~AU. 
This rapid mass loss produces an accretion burst with a duration 
of about 50~yr and magnitude approaching $10^{-4} M_\odot$~yr$^{-1}$. 
During the time period of 650~yr centered on the burst (and outlined by the two
vertical dashed lines), the fragment has lost $24~M_{\rm Jup}$, while only
$8.7~M_{\rm Jup}$ has passed through the sink cell. This means that
only about one third of the mass lost by IF1 passes
through the sink cell triggering the accretion burst, while the remaining two thirds stay 
in the inner disk. In the subsequent evolution, IF1 continues to lose its mass,
but at a slower rate until it finally disperses around $\tilde{t}=3.1-3.3$~yr, producing
another burst of a longer duration and smaller amplitude ($\approx 2\times 10^{-5}~M_\odot$~yr$^{-1}$).
The orbital distance of IF1 ($\approx 25$~AU) does not change notably after the first burst 
and it makes seven revolutions at this quasi-stable orbit\footnote{The animation of clump inward migration can be found at http://www.astro.sfedu.ru/animations/accretion.mp4}.

\begin{figure}
\begin{centering}
\includegraphics[width=1\columnwidth]{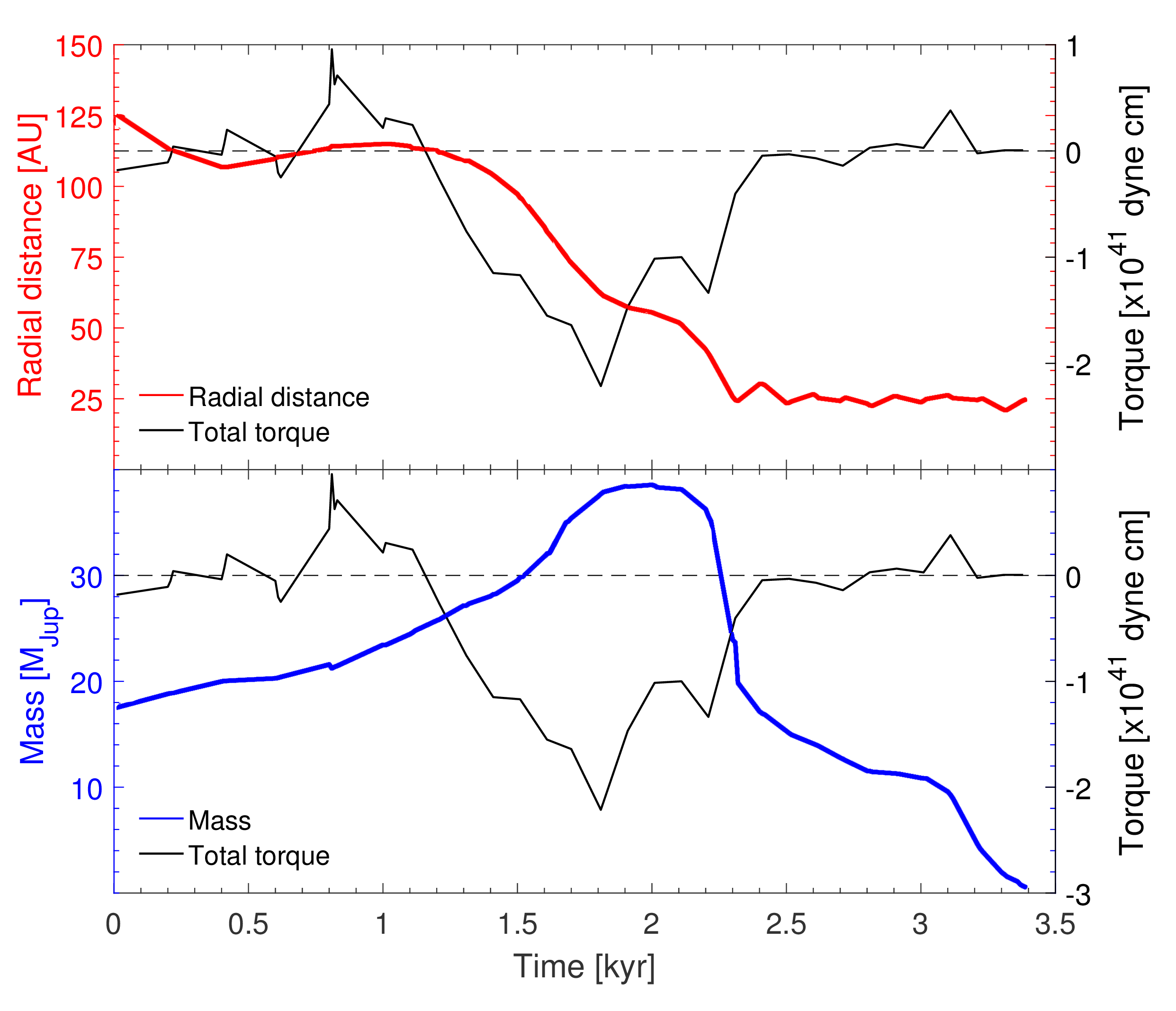}
\par\end{centering}
\caption{\label{fig:11}{\bf Top panel}: Gravitational torque acting
on IF1 from the entire disk including other fragments (the black line) and the radial distance of
IF1 (the red line) as a function of time. {\bf Bottom
panel}: Gravitational torque (the black line) and the mass of IF1 vs. time (the blue line).}
\end{figure}

\subsection{Halt of inward migration}
\label{subsec:Halt-of-migration}
In this section, we discuss the possible reasons for the halt of the IF1 inward migration. 
Figure~\ref{fig:11} shows the total gravitational torque ($\tau_{\mathrm{tot}}$) exerted on 
IF1 by the entire disk (including other fragments), 
the radial distance of IF1 from the protostar, and the mass of 
IF1 as a function of time. Initially, $\tau_{\rm tot}$ stays small, switching in 
sign from positive to negative and back. 
The position of IF1 reflects these changes in $\tau_{\rm tot}$ and IF1 wobbles around a radial 
distance of 110~AU. At $\tilde{t}\eqsim 1.1$~kyr,
a close approach of OF1 with IF1 triggers the fast inward migration of the latter.
Simultaneously, $\tau_{\rm tot}$ becomes negative and increases by absolute value until
$\tilde{t}\eqsim 1.8$~kyr. At this time instance, the process of mass growth reverses and IF1 
starts quickly losing most of its mass. In the subsequent evolution, the total disk torque 
drops (by absolute value) to near zero  and the inward migration of IF1 halts.
We note that the decrease in $\tau_{\rm tot}$ ($\tilde{t}=1.8-2.5$~kyr) correlates with the
burst of accretion, implying that the tidal truncation of IF1 may be the reason
for the ultimate decrease in $\tau_{\rm tot}$.

\begin{figure}
\begin{centering}
\includegraphics[width=1\columnwidth]{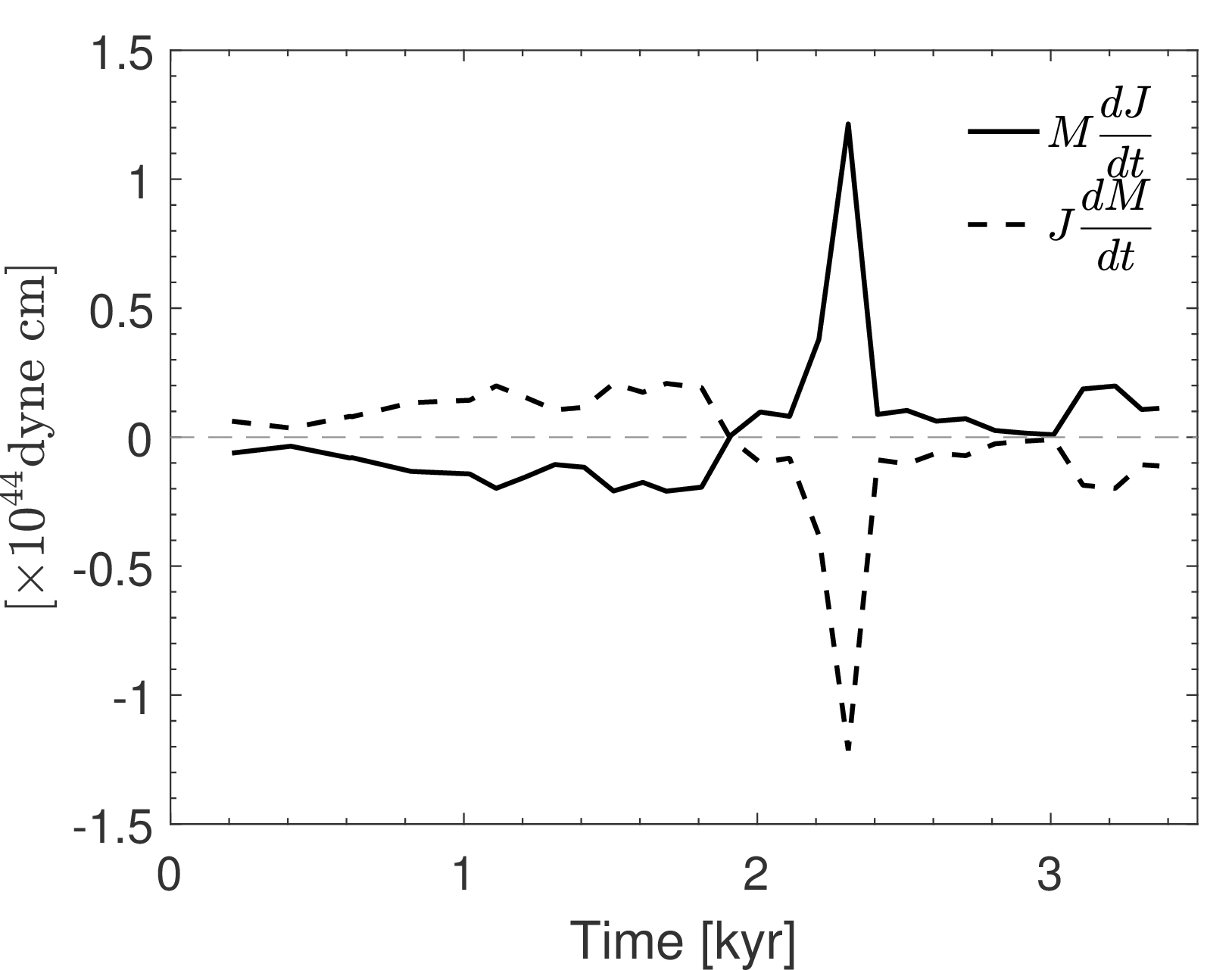}
\par\end{centering}
\caption{\label{fig:12} Time evolution of  $|\boldsymbol{J}|\frac{dM_{\rm f}}{dt}$ (the blue line)  and
$M_{\rm f}\frac{d|\boldsymbol{J}|}{dt}$ (the red line) illustrating the rate of change of the IF1 mass
and specific angular momentum during its migration towards the protostar. The horizontal dashed line
marks the zero line for convenience.}
\end{figure}

To further analyze the inward migration and halt of IF1, we
note that the time derivative of the total angular momentum of IF1 can be expressed as
\begin{equation}
\frac{d\boldsymbol{L}}{dt}=\boldsymbol{J}\frac{dM_{\rm f}}{dt}+M_{\rm f} \frac{d\boldsymbol{J}}{dt},
\label{eq:dL/dt}
\end{equation}
where $\boldsymbol{J=r_{\rm f}\times v}$ is the specific angular momentum of IF1 
and  $\boldsymbol{v}$ is its velocity. Figure~\ref{fig:12} presents
the time evolution of $|\boldsymbol{J}|\frac{dM_{\rm f}}{dt}$ and
$M_{\rm f}\frac{d|\boldsymbol{J}|}{dt}$. Initially, the rate of change of 
$\bl{J}$ is negative and the rate of change of $M_{\rm f}$ is positive,
meaning that IF1 initially gains mass and loses specific angular momentum
when approaching the protostar. After this time instance, however, the signs of 
$|\boldsymbol{J}|\frac{dM_{\rm f}}{dt}$
and  $M_{\rm f}\frac{d|\boldsymbol{J}|}{dt}$ flip and the fragment starts gaining
the specific angular momentum and losing its mass through tidal torques. We note that the specific angular momentum
of IF1 can be related to its centrifugal acceleration as 
\begin{equation}
a_{\rm c.f.} = {|\bl{J}|^2 \over r_{\rm f}^3}. 
\end{equation}
The fact that  $\bl{J}$ increases while $r_{\rm f}$ decreases implies that the centrifugal acceleration
of IF1 quickly increases after $\tilde{t}=1.8$~kyr, which helps to halt its inward migration. We note that both $\boldsymbol{J}dM_{\rm f}/dt$ and
$M_{\rm f}{d\boldsymbol{J}}/{dt}$ are much greater by absolute value than $d \bl{L} / dt$ (see $\tau_{\rm tot}$ in Fig.~\ref{fig:11}), so that small variations in $d \bl{L} / dt$ during the inward migration of IF1 do not invalidate our analysis.
We conclude that the tidal truncation of IF1 and the associated increase in its specific angular momentum
helps to halt the fast inward migration of IF1.

\subsection{Formation of a protoplanetary core}
The fragments considered in our numerical simulations are the first hydrostatic cores in the parlance
of star formation. Their sizes range between a few AU and a few tens of AU, which makes them vulnerable
to the action of stellar tidal torques at radial distances of several tens of AU and smaller. 
However, when the gas temperature in the interiors
of the fragment exceeds $\approx 2000$~K, molecular hydrogen dissociates and 
the so-called second collapse ensues, leading to the formation of a protoplanet or a proto-brown dwarf,
which can withstand the stellar tidal torques even at sub-AU distances. The second collapse
is difficult to resolve in global disk simulations as our own. Nevertheless, we can estimate 
which part of IF1 would collapse to form a planetary-sized core and which part would
form a circumplanetary disk and/or an envelope. We note that IF1 is characterized by 
fairly strong rotation before the supposed onset of the second collapse, with the ratio
of rotational to gravitational energy $\beta_{IF1}=9.5\%$. 
Therefore, we may expect the formation of a fairly massive
disk and/or envelope during the second collapse.

\begin{figure}
\begin{centering}
\includegraphics[width=1\columnwidth]{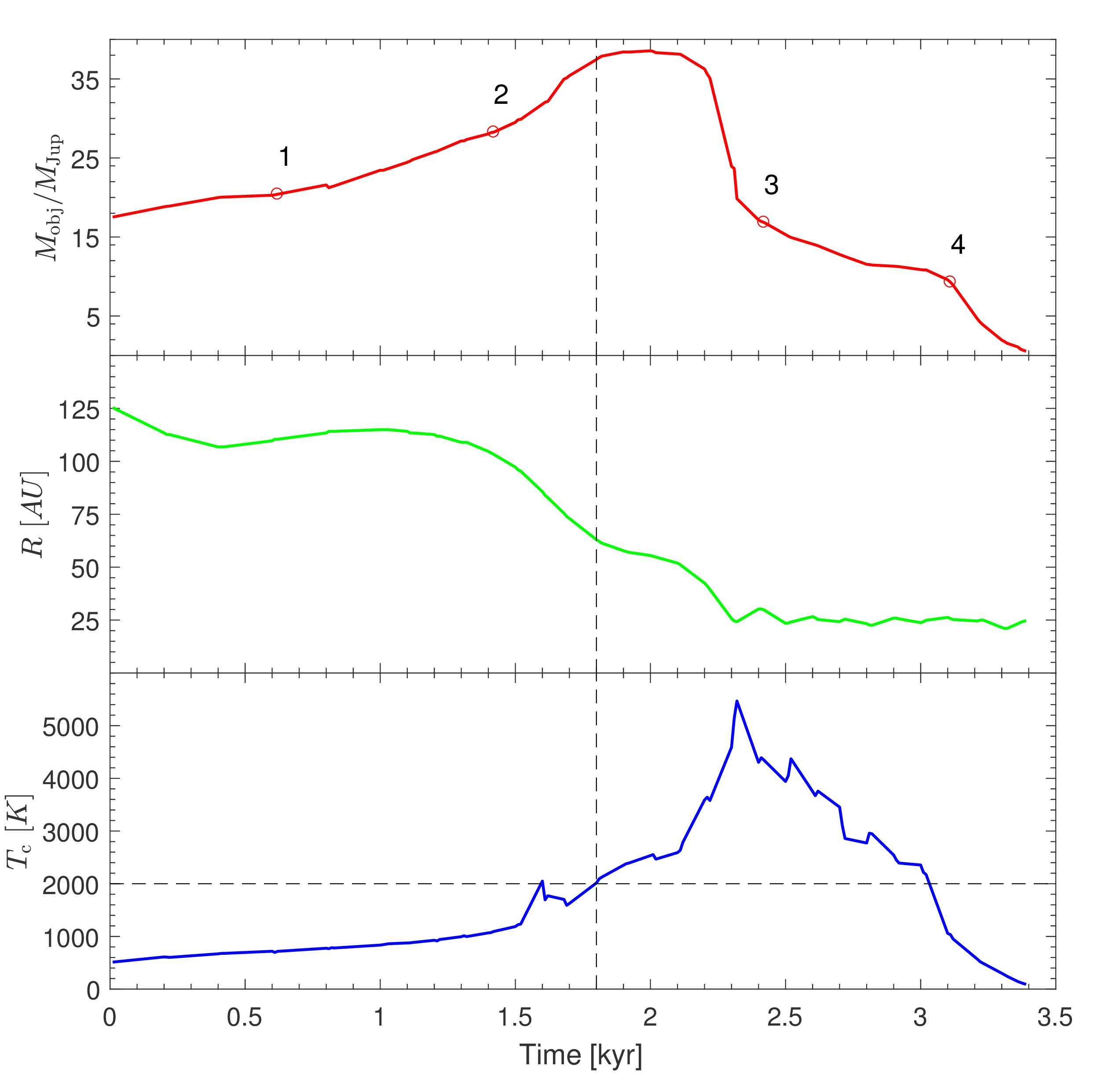}
\par\end{centering}
\caption{\label{fig:13} Mass of IF1 vs. time (top panel), 
radial distance of IF1 vs. time (middle panel), and central temperature of IF1 vs. 
time (bottom panel). The horizontal
dashed line shows a threshold temperature of 2000~K, above which molecular
hydrogen dissociates and the second collapse is supposed to ensure. 
The vertical dashed line shows the time
instance when the central temperature of IF1 reaches the threshold value.
The red circles indicate time instances at which the mass growth rate of IF1 is calculated: 1) $6.6 \times 10^{-3} M_{\rm Jup}$~yr$^{-1}$, 2) $1.45 \times 10^{-2} M_{\rm Jup}$~yr$^{-1}$, 3) $-1.8 \times 10^{-2} M_{\rm Jup}$~yr$^{-1}$, and 4) $-3.3 \times 10^{-2} M_{\rm Jup}$~yr$^{-1}$.   }
\end{figure}

Figure~\ref{fig:13} presents the mass, radial distance, and 
central temperature of IF1 as a function of time. The vertical dashed line
indicates the time instance ($\tilde{t}=1.8$~kyr)  when the central temperature reaches 2000~K 
and the second collapse is supposed to take place. This is also the time when the mass of IF1
reaches a maximum value of about $38~M_{\rm Jup}$.  The central temperature
continues to increase until IF1 is heavily truncated by the tidal torques at $\tilde{t}=2.3$~kyr.
In the subsequent evolution, the central temperature drops again.

\begin{figure}
\begin{centering}
\includegraphics[width=1\columnwidth]{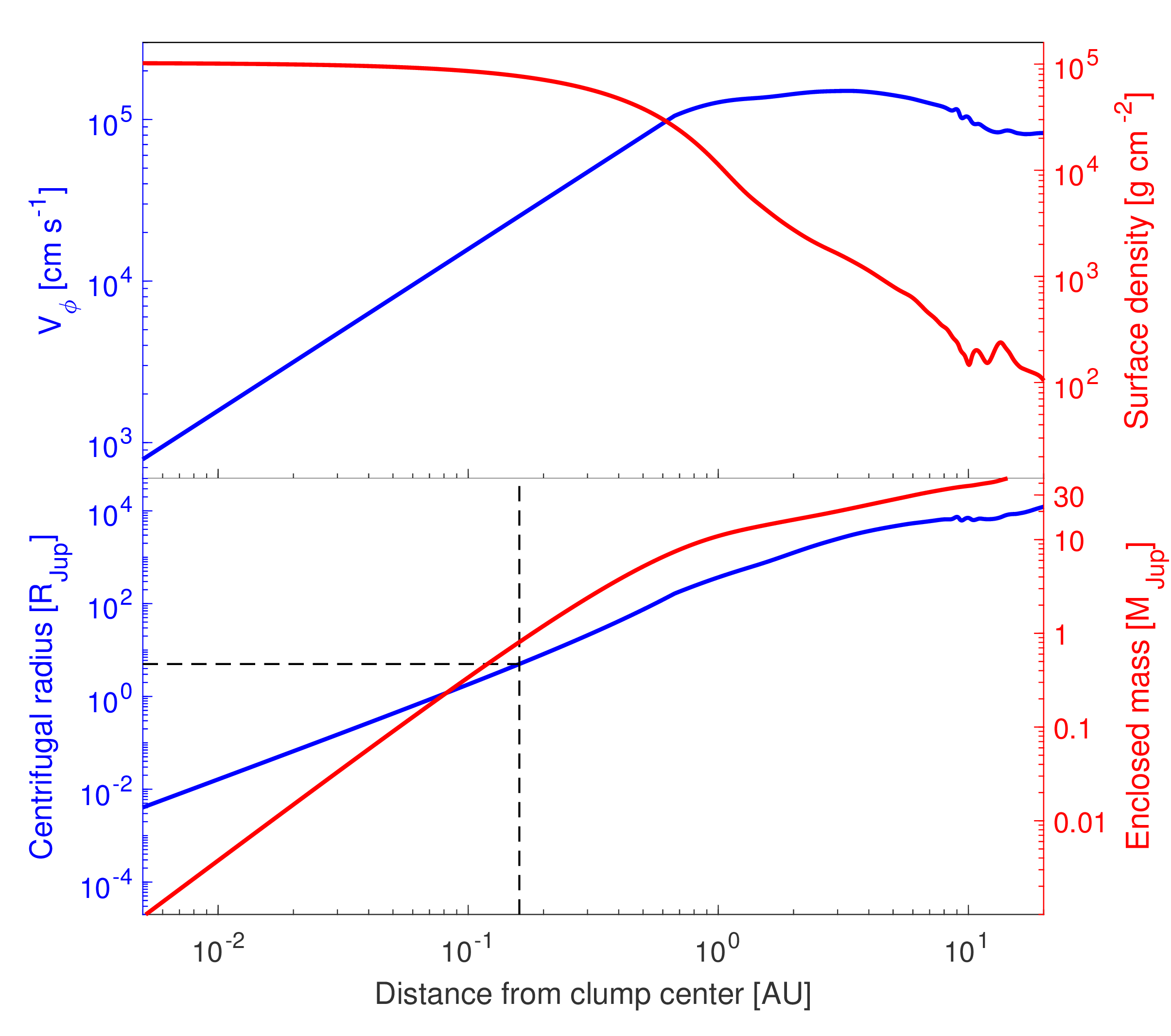}
\par\end{centering}
\caption{\label{fig:11a}{\bf Top panel}. Azimuthally averaged surface density and angular velocity profiles of IF1 as a function of the radial distance from the center
of the fragment. {\bf Bottom panel}. The centrifugal radius and the enclosed
mass of IF1 as a function of the radial distance from the center of
the fragment. The horizontal dashed line marks the radius of 
the second hydrostatic core $R\mathrm{_{s.core}}=5R_{\mathrm{Jup}}$. The vertical dashed line separates the inner and outer parts of IF1 that form a protoplanetary core and disk plus envelope, respectively, during the second collapse caused by $H_2$ dissociation.}
\end{figure}

To calculate the fraction of IF1 that would form the protoplanetary core at $\tilde{t}=1.8$~kyr, we split IF1 into 
a series of concentric annuli and calculated the azimuthally averaged surface density and angular velocity profiles (in the frame of reference of the center of IF1) as a function of the radial distance from the center of the fragment. The first data point lies at r=0.5~AU and we extrapolated both distributions to smaller radii assuming that $v_{\phi}$ approaches zero in the center (no gravity point-source) and $\Sigma$ has a constant density plateau in the inner regions of IF1. The corresponding profiles are shown in the top panel of Figure~\ref{fig:11a}. We then calculated the centrifugal 
radius of each annulus using the following expression
\begin{equation}
R_{\mathrm{cf}}= {|\bl{J}|^2 \over G M_{\rm f}(r)},
\label{centrad}
\end{equation}
where $M_{\rm f}(r)$ is the mass contained within the radial distance $r$ from the center of IF1.
Equation~(\ref{centrad}) assumes that the specific angular momentum $\bl{J}$ of each annulus is conserved during the second collapse. 
%We calculated $|\bl{J}|$ using the azimuthally averaged
%angular velocities in the frame of reference of the center of IF1.

The bottom panel in Figure~\ref{fig:11a} presents the centrifugal radius and the enclosed mass
of IF1 as a function of the radial distance from the center of the fragment. The horizontal dashed line marks the radius of the second hydrostatic core, 
taken to be $R_{\rm s.core}=5 R_{\rm Jup}$. We note that the exact value of $R_{\rm s.core}$ 
is rather uncertain, but most studies assume it to vary from several to ten $R_{\rm Jup}$ 
\citep{2012BaraffeVorobyov,2011HosokawaOffner,2017VorobyovElbakyan}.
Part of IF1 that lies to the left of the vertical dashed line at $R_{\rm crit}=0.16$~AU is characterized by $R_{\rm cf}\le R_{\rm s.core}$. During the second collapse, all material that lies inside $R_{\rm crit}$  would directly form the protoplanetary core. 
On the other hand, the material that lies outside $R_{\rm crit}$
would hit the centrifugal barrier before landing on the protoplanetary core and would 
rather form a disk and/or envelope around the core.

We found that if IF1 had collapsed at $\tilde{t}=1.8$~kyr, it would have formed a protoplanetary core 
with a mass of $0.81~M_{\rm Jup}$, which is only a small fraction of the fragment's total mass at  this time instance ($37~M_{\rm Jup}$). 
If we increase or decrease the radius of the second hydrostatic core by a factor of 2, the resulting mass of the protoplanetary core becomes $1.4~M_{\rm Jup}$ or $0.45~M_{\rm Jup}$, respectively.
 We note that the core will most certainly continue growing in mass through accretion, but the terminal mass is uncertain.
We expect that during the subsequent inward migration, the protoplanetary core may lose part of its
circumplanetary material through tidal torques, perhaps creating an accretion burst similar to what has happened with IF1. 
The protoplanetary core itself would, however, survive and settle at an orbit of about a few tens of AU. 

In a future work, we plan to test this hypothesis introducing a sink particle 
at the time instance of supposed second collapse.
A number of young stars have planetary-mass companions orbiting the
host star at radial distances from 10 to 100 AU (e.g., Fomalhaut~b, 51~Eri~b,
HR~8799). For instance, four planets of HR~8799 (b, c, d, e) orbit the star at radial distances of 
15, 24, 38, and 68~AU, respectively. The second collapse of IF1 followed 
by tidal truncation 
could explain the formation of planets b, c, or d in the HR~8799 system.

\section{Effects of the inner boundary and stellar motion}
\label{effects}

%\subsection*{\textit{Computational grid is interpolated to make possible the calculation
%of the mass of the hydrogen core.}}

%\subsection{WOBBLING AND BOUNDARY }

In the fiducial model used in the previous sections we imposed a free outflow
inner boundary condition, so that the matter was allowed to flow
from the computational domain to the sink cell, but was prevented from flowing 
from the sink cell back to the computational domain. We also set a fixed star
in the coordinate center and did not allow it to move in response to the 
gravity force of the disk and fragments. In this section, we relax both assumptions. 
We consider two models: one with the free inflow-outflow boundary condition (see Sect.~\ref{sec:modify}),
but the fixed star (hereafter, model~IOB, "inflow-outflow-boundary") and the other with 
the free outflow boundary condition, but with the star moving in response to the gravity force 
of the disk (hereafter, model~SM, "stellar motion"). 
Stellar motion may be important when the disk is massive and  strongly asymmetric, as is the case 
for gravitationally unstable and fragmenting disks. In this case, the rotation will be around the 
center of mass of the system, rather than around the central star itself.
We restarted our fiducial model from $\tilde{t}=-3.5$~kyr, but with the aforementioned modifications.

\begin{figure}
\begin{centering}
\includegraphics[width=1\columnwidth]{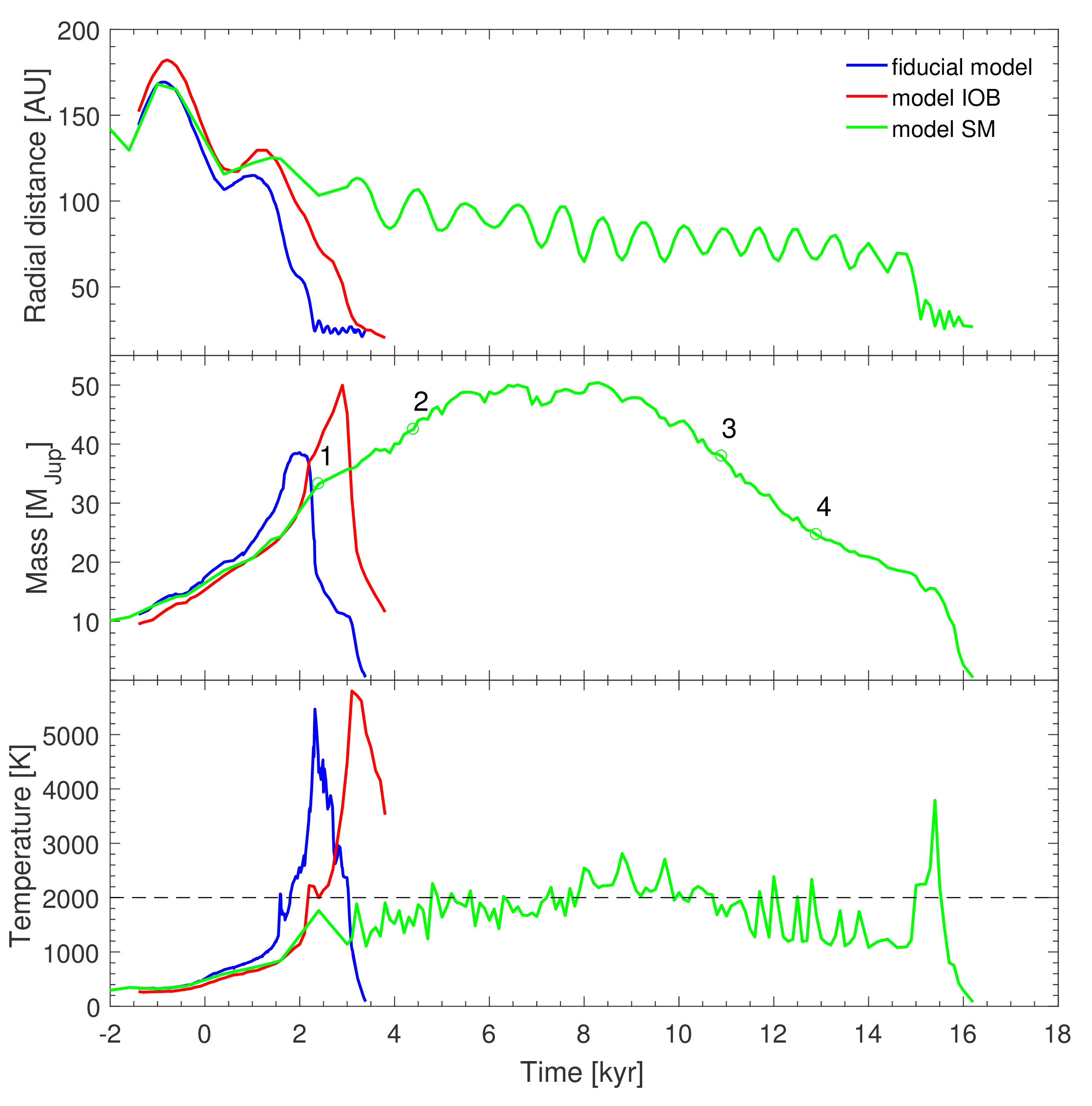}
\par\end{centering}
\caption{\label{fig:16}Radial distance (top panel), mass (middle panel), and
central temperature (bottom panel) of fragment IF1 vs. time in the fiducial model (blue curves),
IOB model (red curves) and SM model (green curves).  The green circles indicate time instances at which the mass growth rate of IF1 is calculated: 1) $4.0 \times 10^{-3} M_{\rm Jup}$~yr$^{-1}$, 2) $4.8 \times 10^{-3} M_{\rm Jup}$~yr$^{-1}$, 3) $-3.5 \times 10^{-3} M_{\rm Jup}$~yr$^{-1}$, and 4) $-2.6 \times 10^{-3} M_{\rm Jup}$~yr$^{-1}$.   }
\end{figure}

\begin{figure}
\begin{centering}
\includegraphics[width=1\columnwidth]{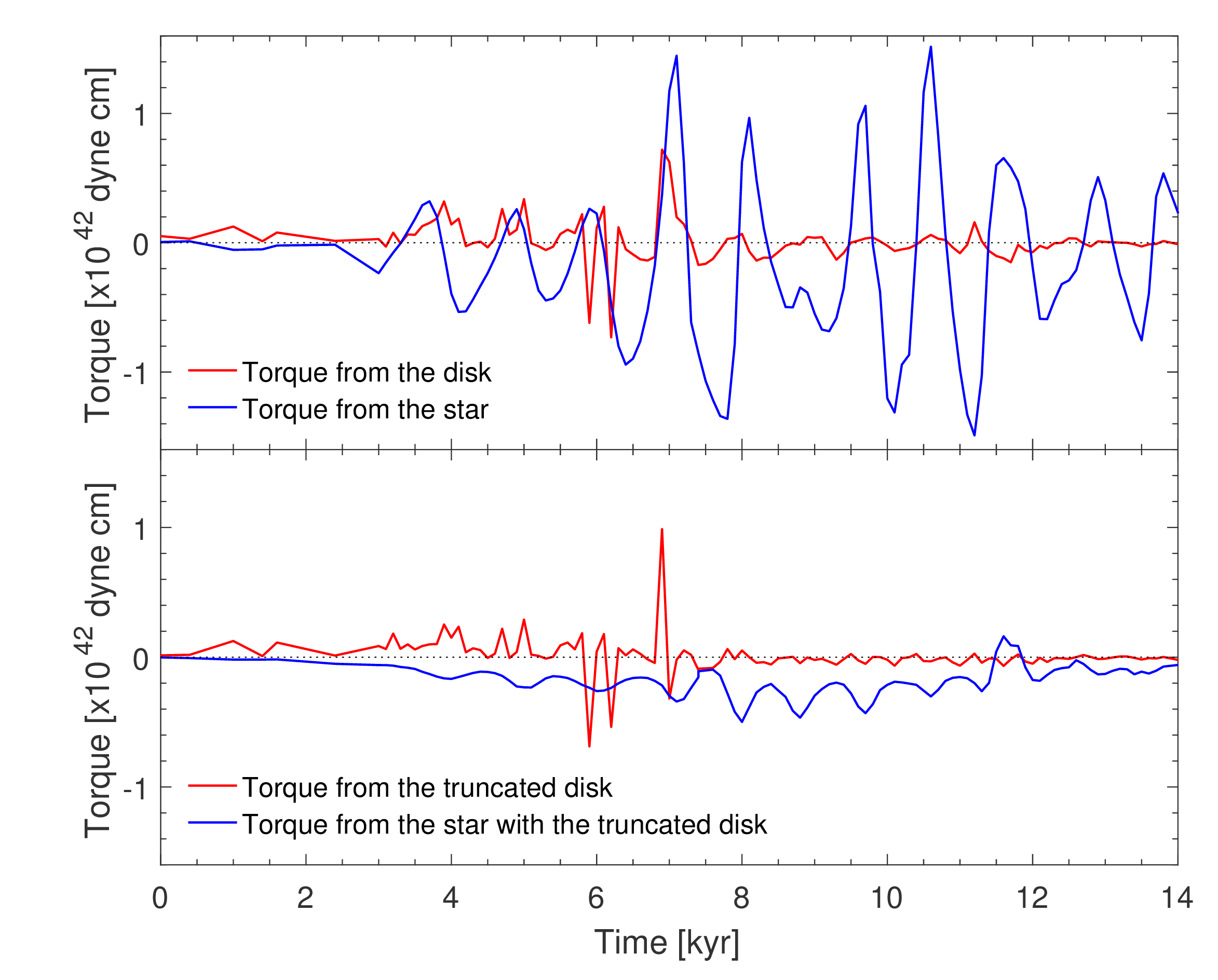}
\par\end{centering}
\caption{\label{fig:15}{\bf Top panel}. Torques exerted on IF1 by  the entire disk (the red solid line) and by the central star (the blue solid line). {\bf Bottom panel}. The truncated disk and stellar torques exerted on IF1 and calculated using only the material lying in the inner 145~AU of the disk. }
\end{figure}

Figure \ref{fig:16} presents the radial distance (top panel), mass
(middle panel), and central temperature (bottom panel) of IF1 
during its inward migration for the fiducial model, model~IOB, and model~SM. 
Clearly, the general evolution is similar in all three models -- IF1 migrates
inward, first gaining and then loosing its mass, until it finally disperses through
the action of tidal torques at a radial distance of 25--30~AU.
However, there are some differences in the details of inward migration that we describe
below.

The inner boundary condition has a moderate effect on the dynamics
and properties of IF1. The migration timescale of IF1 in model~IOB
is slightly longer and its maximum mass is slightly higher than in the fiducial model. The inflow-outflow boundary condition reduces 
the artificial drop in the gas surface density near 
the inner edge of the disk. As a result, there is more material
in the inner disk interior of the IF1 orbit  
and this material exerts a positive gravitational torque on IF1, slowing down its inward migration.

\begin{figure*}
\begin{centering}
\includegraphics[width=2\columnwidth]{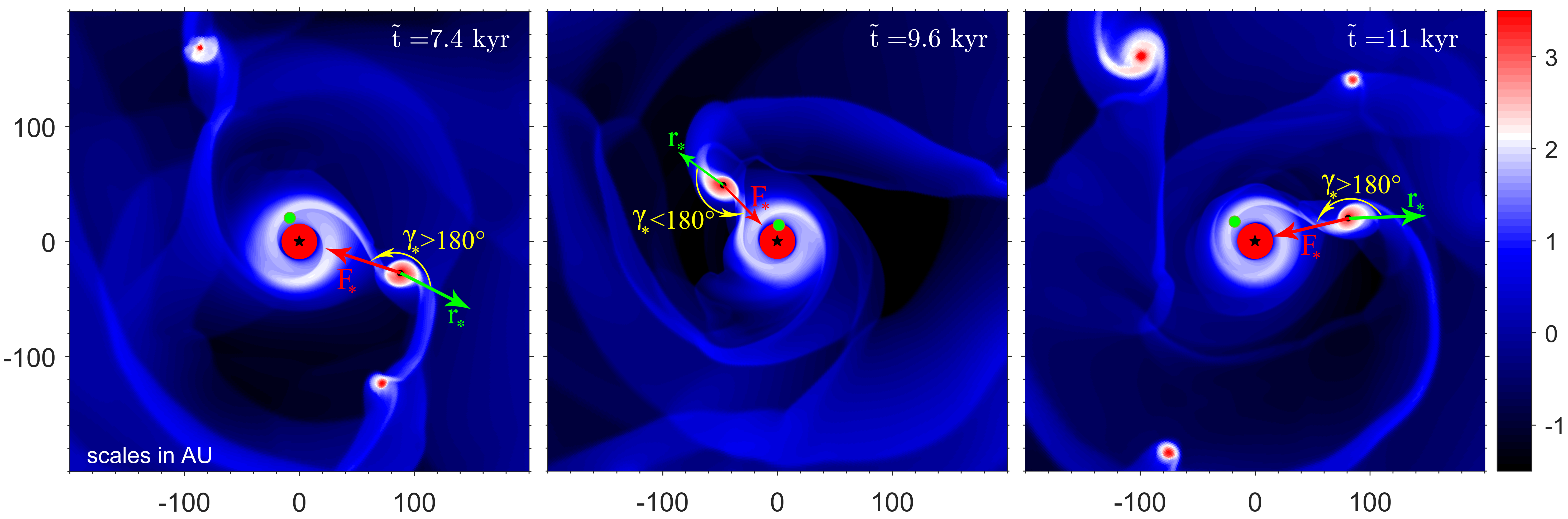}
\par\end{centering}
\caption{\label{fig:14}Gas surface density distributions in the SM model at several time instances focusing on IF1. The direction of the gravity force $F_\ast$ acting on IF1 from the star is marked with the red
arrow. The direction of the lever arm $r_\ast$ used for the calculation
of the torque acting from the star on IF1 is marked with the green arrows. The curved yellow arrows show the
angle $\gamma_\ast$ between $r_\ast$ and ${F_\ast}$. The disk rotates counter-clockwise. The scale bar is in log g~cm$^{-2}$.}
\end{figure*}

On the other hand, allowing for stellar motion significantly changes 
the dynamics and properties of IF1. The timescale of inward migration in model~SM is much longer than in the other two models without stellar motion. 
The slowed-down migration in the presence of stellar motion was previously reported in the context of planet migration by, e.g., \citet{2017RegalyVorobyova}. To explore the reason for the slowed-down migration, we calculated the gravitational torques exerted on IF1 by the entire disk and by the central star with respect to the center of mass of the entire disk plus star system, and plotted these values as a function of time in the top panel of Figure~\ref{fig:15}. Clearly, the stellar torque is greater by the absolute value than the disk torque thanks to the greater mass of the star ($M_\ast=0.51~M_\odot$ vs. $M_{\rm disk}=0.18~M_\odot$) and proximity of the star to IF1 (resulting in a stronger gravity force). Moreover, both torques show an oscillatory behavior, changing their signs from positive to negative and back.

The alternating sign of the stellar torque is illustrated in Figure~\ref{fig:14} showing the gas surface density distributions in
model~SM at several time instances.  The center of mass of the disk plus star system is marked by the green dots and the position of the central star is shown by the black star symbol in the coordinate center. 
The red arrows show the direction of gravitational force 
$F_\ast$ acting on IF1 from the star, while the green arrows show the direction of the lever arm $r_{\rm c.m.}$ (with respect to the center of mass) used in the calculation of the torque exerted on IF1 by the star. The yellow 
arrows show the angle $\gamma_\ast$ between $r_{\rm c.m.}$ and $F_\ast$.  The sign of the torque $\tau_\ast= r_{\rm c.m.} \, F_\ast \, \sin\gamma_\ast$ exerted on IF1 by the star depends on the value of the angle $\gamma_\ast$: the torque
is positive when $\gamma_\ast<180^\circ$ and negative when $\gamma_\ast>180^\circ$. 
The angle $\gamma_\ast$ in its turn depends on the spatial arrangement of IF1, the star, and the center of mass.  When the center of mass is leading the fragment,  $\gamma_\ast>180^\circ$ and vice versa (note that the disk and fragment rotate counterclockwise).

The alternating sign of both torques affect the character of IF1 migration - it shows alternating inward-outward short-amplitude excursions, but the net result is a slow inward migration, in agreement with the integrated (disk plus star) torque being negative in sign, $-1.88\times 10^{43}$~dyne~cm, bu the end of migration. 
The slowed-down inward migration also changes the internal properties of IF1.
In the model with stellar motion, IF1 accumulates and loses mass much slower than in models without stellar motion (middle panel in Fig.~\ref{fig:16}),
although the maximum attainable mass of IF1 is similar to the other two models.
As a result of slow mass accumulation in model~SM, the gas temperature in the
center of IF1 also grows slowly (bottom panel in Fig.~\ref{fig:16}). The
maximum attainable temperature is also lower ($\approx 3800$~K) than in the
other two models ($\approx 5500$~K). 

The center of mass of the system and, hence, the disk and stellar torques are determined by the global disk structure and position of other fragments in the disk. To illustrate this, we plot in the bottom panel of Figure~\ref{fig:15} the disk and stellar torques, but calculated using the material in the inner 145~AU,
so that only IF1, the inner part of the disk, and the central star are considered. Clearly, the stellar torque is reduced significantly and it is now mostly negative, meaning that the character of IF1 migration would be different in this case - it would likely be a steady inward migration at a faster speed.

\begin{figure*}
\begin{centering}
\includegraphics[width=2\columnwidth]{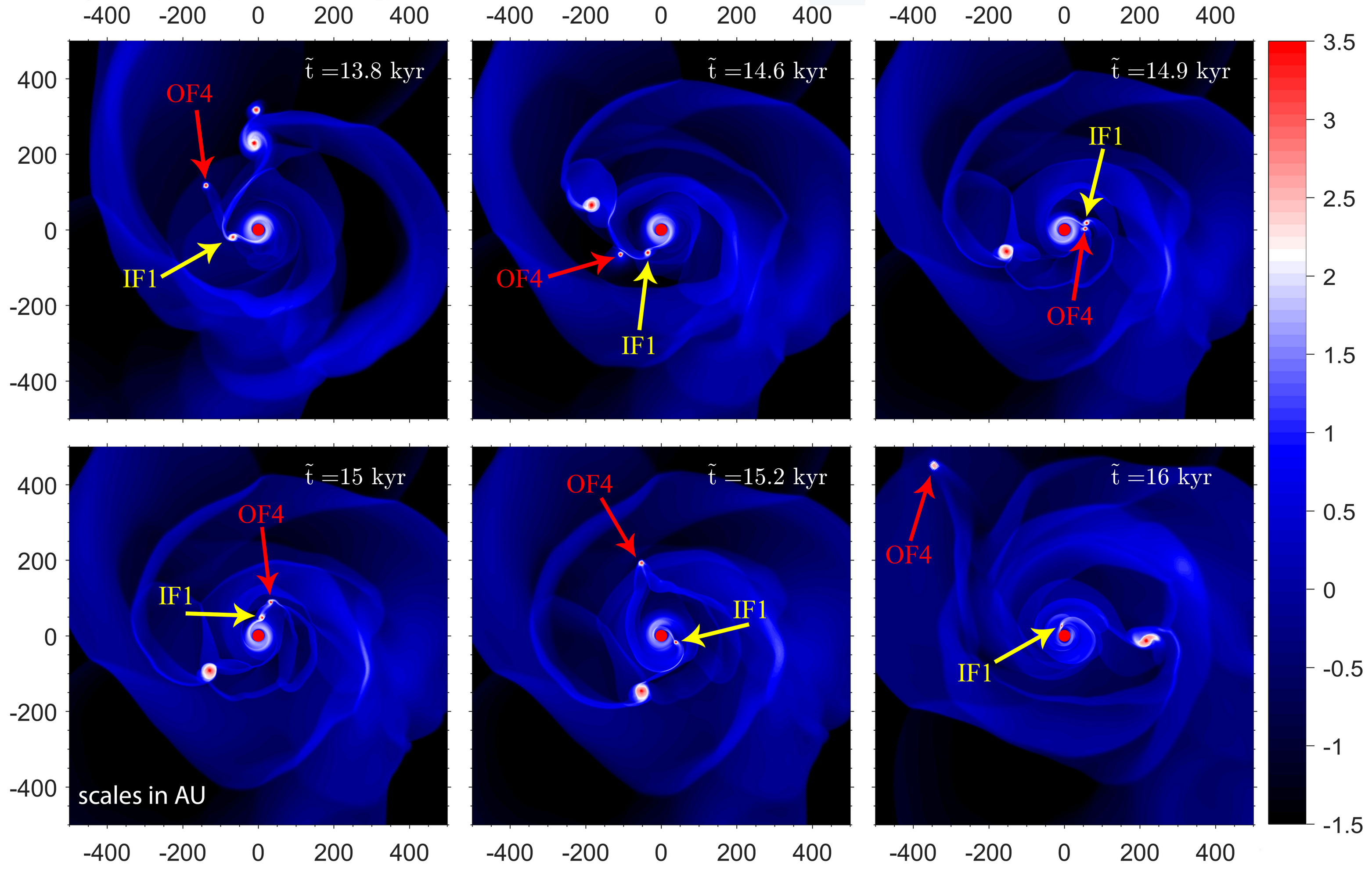}
\par\end{centering}
\caption{\label{fig:18}Gas surface density distributions in the SM model at
several time instances focusing on the close encounter between IF1 and OF4,  resulting in the ejection of the latter fragment. IF1 and OF4 are indicated with the yellow and red arrows, respectively. The disk rotates
counter-clockwise.}
\end{figure*}

\begin{figure}
\begin{centering}
\includegraphics[width=1\columnwidth]{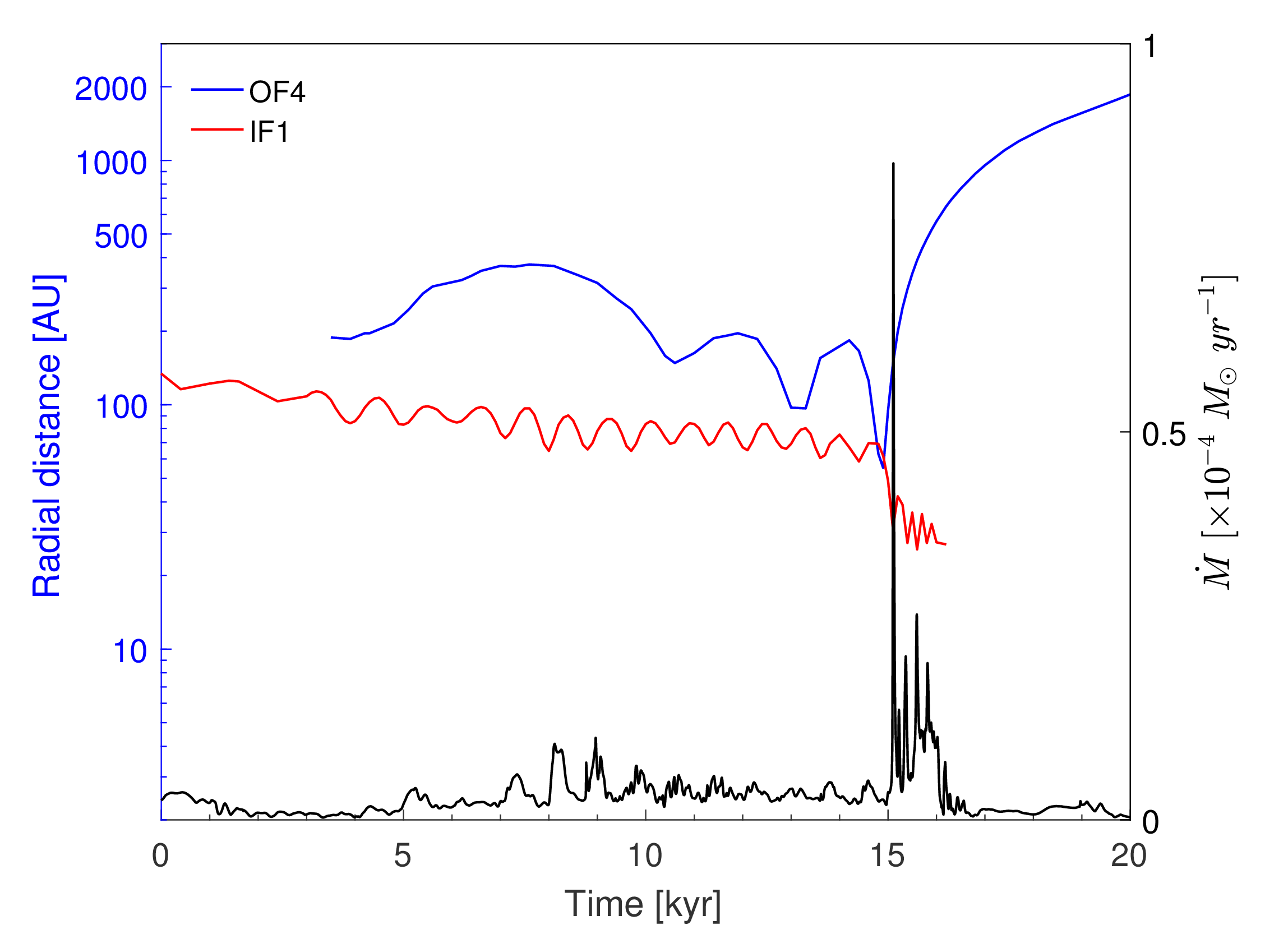}
\par\end{centering}
\caption{\label{fig:17} Radial distances of IF1 and OF4 from the star (the red and blue lines, respectively, together with protostellar accretion rate (the black line), as a function of time in model SM. }
\end{figure}

Model SM at $\tilde{t}\approx 15$~kyr  shows an interesting phenomenon -- the radial distance and the mass of IF1 
sharply decrease, while its central temperature increases. During the subsequent 1.0~kyr of evolution the fragment disperses. This is caused by a close encounter of IF1 with another fragment (OF4 - outer fragment 4) shown in Figure~\ref{fig:18} with the red arrows. The closest approach occurs at $\tilde{t}\approx14.9$~kyr and leads to the ejection of OF4 due to the multi-body gravitational interaction. The ejected fragment has a mass of 7.5~$M_{\rm Jup}$ and a velocity of 1.8~km~s$^{-1}$. The escape speed of the star plus disk system, $v_{\rm esc}=1.2$~km~s$^{-1}$, is smaller than the velocity of OF4, meaning that we witness a true ejection and not the scattering of a planetary mass object to a wider orbit. The close approach is a paired effect -- the ejection of OF4 is causing IF1 to quickly lose its angular momentum and approach the star\footnote{The animation of clump ejection can be found at http://www.astro.sfedu.ru/animations/intruder.mp4}. The approach is accompanied by tidal destruction and an accretion burst consisting of one strong peak and several smaller ones, as is shown by the black line in Figure~\ref{fig:17}. This Figure also shows the radial distances of OF4 and IF1 from the star. Clearly, OF4 moves on a highly eccentric orbit caused by the gravitational interaction with other fragments in the outer disk. During one of its close approaches to the star, the trajectory of OF4 intersects with that of IF1, leading to the ejection of OF4 and accelerated infall of IF1.  
This phenomenon is therefore a chance effect, but its probability is enhanced by slowed-down  migration of IF1 and eccentric orbit of OF4.

Interestingly, the close approach between OF4 and IF1 causes a sharp increase in the temperature of the latter, probably due to tidal heating. The temperature exceeds 2000~K, which indicates that close approaches between the fragments can facilitate the second collapse and formation of protoplanetary cores.

 Finally, it is interesting to compare the results of our simulations with the recent works of \citet{2017Nayakshinb} and \citet{2018StamatellosInutsuka} on the dynamics of protoplanets in gravitationally unstable disks. The work of Nayakshin is most relevant to our study as it also considers the dynamics of gaseous clumps rather than collapsed sink particles. He found that 
the dynamics of the clumps depends on the dust opacity (scaled up or down with respect to that of \citet{2009ZhuHartmann}), on the initial seed mass of the clump, and even on the initial azimuthal position of the clump in the disk. Depending on these parameters, the clumps may either quickly migrate towards the star and disperse or quickly gain mass and effectively halt their inward migration at tens of au. 
The migration timescales are similar to what was found in our models for the case of a motionless central star and the maximum attainable clump masses, a few tens of Jupiter masses, are also in agreement with our study. There is however a significant difference in the initial setup of the disk. Nayakshin considers a disk that has only one fragment at a time, while our disks are heavily fragmented and have several fragments at a time. The clump-to-clump interaction introduces another degree of complexity in the clump dynamics and evolution, leading in our models to inward migration of clumps which are otherwise quasi-stable in the Nayakshin models. 

It is less straightforward to compare our models with those of \citet{2018StamatellosInutsuka}, because they consider the dynamics of point-sized sink particles, i.e., the objects that have already experienced second collapse to planetary densities due to dissociation of molecular hydrogen,
while we consider the pre-collapsed gaseous clumps.
Nevertheless, the estimated mass growth rates of our clumps in the model with stellar motion (see Fig.~13) are in good agreement with 
a mass accretion rate on the protoplanet, $\sim 10^{-3} M_{\rm Jup}$~yr$^{-1}$, found in their simulations. The model without stellar motion (Fig. 11), however, yields mass growth rates that are an order of magnitude higher than what is found in \citet{2018StamatellosInutsuka}.
The effect of protolanetary radiative feedback, which was found to be important for setting the final mass of the protoplanet, can only be studied once we introduce sink particles in our models.

\section{Conclusions}
\label{conclude}

We used high-resolution grid-based numerical hydrodynamics simulations of disk formation and evolution to study the migration of dense gaseous clumps that form in the disk through gravitational fragmentation. Our numerical simulations cover the entire embedded phase of disk evolution, starting from the collapse of a prestellar core and ending with its complete dissipation due to accretion on the star plus disk system.  Thanks to the logarithmically spaced grid in the radial direction we achieved a sub-AU resolution in the disk regions where fragmentation and migration takes place, which allowed us to study the internal structure of migrating clumps in detail. Our findings can be summarized as follows.
\begin{itemize}

\item Gaseous clumps that form in the outer disk regions are often perturbed by other clumps or disk structures, such as spiral arms, and migrate toward the central star. When approaching the star, the clumps lose most of their
diffuse envelopes  through tidal torques. 
The tidal mass loss helps the clumps to significantly slow down or even halt 
their inward migration at a distance of a few tens of AU from the protostar.

\item 
Tidal truncation of gaseous clumps as they approach the protostar can produce accretion bursts similar in magnitude and duration to the FU-Orionis-type eruptions, if the tidally stripped material is  accreted by the protostar. Numerical simulations with a smaller inner sink cell (15~AU in the current study) are needed to further investigate this phenomenon. 

\item
Tidal truncation and associated halt of inward migration produce hot and dense gaseous nuclei at distances on the order of a few tens of AU. These nuclei may further experience the second collapse down to planetary densities through the dissociation of molecular hydrogen at $T>2000$~K. However, only a small fraction of the total clump mass is expected to directly form the protoplanetary core and most of the clump material would form the circumplanetary disk and/or envelope.

\item
The details of inner boundary implementation have a minor effect on the properties and migration of gaseous clumps, whereas stellar motion can significantly increase their inward migration timescale. The slowed-down migration increases the probability of chance encounters with other eccentric-orbit clumps, leading in some cases to the ejection of the least massive (planetary-mass) objects from the disk in the interstellar medium. 

\end{itemize}

We conclude that tidal truncation of gaseous clumps is an important effect that not only slows down (or even halts) their
inward migration, but also facilitates the formation of giant protoplanets on tens of AU orbits (as also predicted by tidal downsizing theory, see  \citet{2017Nayakshin}) and triggers accretion bursts, which in turn may affect the dust growth, chemistry, and gravitational fragmentation in protostellar disks \citep[e.g.,][]{2013VorobyovBaraffe,2015Stamatellos,2017Hubbard,2017RabElbakyan}. A premature introduction of sink particles as proxies for gaseous clumps carries a risk of neglecting these important phenomena.

\section*{Acknowledgements}
We are thankful to the anonymous referee for useful comments and suggestions which helped to improve the manuscript. We are thankful to Nader Haghighipour for stimulating discussions.
This work was supported by the Russian Science Foundation grant 17-12-01168. The simulations were performed on the Vienna Scientific 
Clusterand on the Compute Canada network

\bibliographystyle{aa}
\bibliography{ref_clumps}
\end{document}